\documentclass[10pt, conference, compsocconf]{IEEEtran}
\ifCLASSINFOpdf
   \usepackage[pdftex]{graphicx}
\else
   \usepackage[dvips]{graphicx}
\fi

\usepackage{comment}

\usepackage[tight,footnotesize]{subfigure}
\begin{document}
\title{Diffusion in Networks With Overlapping Community Structure}

\author{\IEEEauthorblockN{Fergal Reid}
\IEEEauthorblockA{Clique Research Cluster\\
University College Dublin\\
Email: fergal.reid@gmail.com}
\and
\IEEEauthorblockN{Neil Hurley}
\IEEEauthorblockA{Clique Research Cluster\\
University College Dublin\\
Dublin 4, Ireland\\
}

}

\maketitle

\begin{abstract}
In this work we study diffusion in networks with community structure.
We first replicate and extend work on networks with non-overlapping community structure.
We then study diffusion on network models that have overlapping community structure.
We study contagions in the standard SIR model, and complex contagions thought to be better models of some social diffusion processes.
Finally, we investigate diffusion on empirical networks with known overlapping community structure, by analysing the structure of such networks, and by simulating contagion on them.
We find that simple and complex contagions can spread fast in networks with overlapping community structure.
We also find that short paths exist through overlapping community structure on empirical networks.
\end{abstract}

\section{Introduction}
Diffusion on complex networks is often of interest: we may be investigating how quickly a piece of news can travel through a social network; whether a virus can spread through a computer network; or if a disease will become a pandemic.
All these diffusion processes are influenced by the topology of the network on which they occur.
Early work on the idea of `six degrees of separation' by Milgram \cite{milgram1967small} and the later analytical work of Watts and Strogatz \cite{watts1998collective} informed our intuition about the `small world' of the human social network, and the speed with which a contagion may traverse a large network.

It has been established in the epidemiological literature that the structure of networks influences diffusion on them \cite{aral1999sexual, helleringer2007sexual}.
Much work has been done on modelling and analysis of simple contagions on Watts-Strogatz networks, and on `scale free' networks generated from the Barabasi-Albert model \cite{pastor2001epidemic,Newman2002a, Meloni2009}.
However, the Watts-Strogatz model, though clustered, is not a good model of real social networks, with all nodes connected in a giant ring lattice.
The Barabasi-Albert model, on the other hand, does not generate networks which are highly clustered.
Real social networks are known to have community structure and many algorithms have been developed to uncover this structure \cite{fortunato2009community}.
Synthetic network models have been proposed which capture community structure in various forms \cite{guillaume2006bipartite, botha2010community,newman2001random, lancichinetti2009benchmarks}. 
However, the effects of community structure on diffusion are still under investigation \cite{liu2005epidemic, chu2009epidemic, salathe2010dynamics}.
While there is some variation between the community network models these authors study diffusion on, in all cases communities are modelled as non-overlapping dense subgraphs, with weak ties randomly inserted between them.
In particular, Salath\'{e} and Jones \cite{salathe2010dynamics} use SIR simulation to investigate epidemic spread on synthetic networks, generating their networks by creating a dense Watts-Strogatz graph for each community, and then adding random edges between these communities.
However, such models, where well defined communities are connected by narrow `bridges' of randomly inserted `weak ties', are not descriptive of the overlapping community structure known to be found in real social networks. In fact, Leskovec et al. \cite{leskovec2008statistical} showed that well defined, `non-overlapping' communities -- groups of nodes with small \textit{conductance} -- do not exist at large scales in many empirical networks. In \cite{reid2011partitioning} it is established that communities overlap pervasively in a range of empirical networks, and that it is not possible to partition many of these networks without splitting communities.
\begin{figure*}[!htb]
\begin{center}
    \subfigure[]{
        \includegraphics[width=40mm]{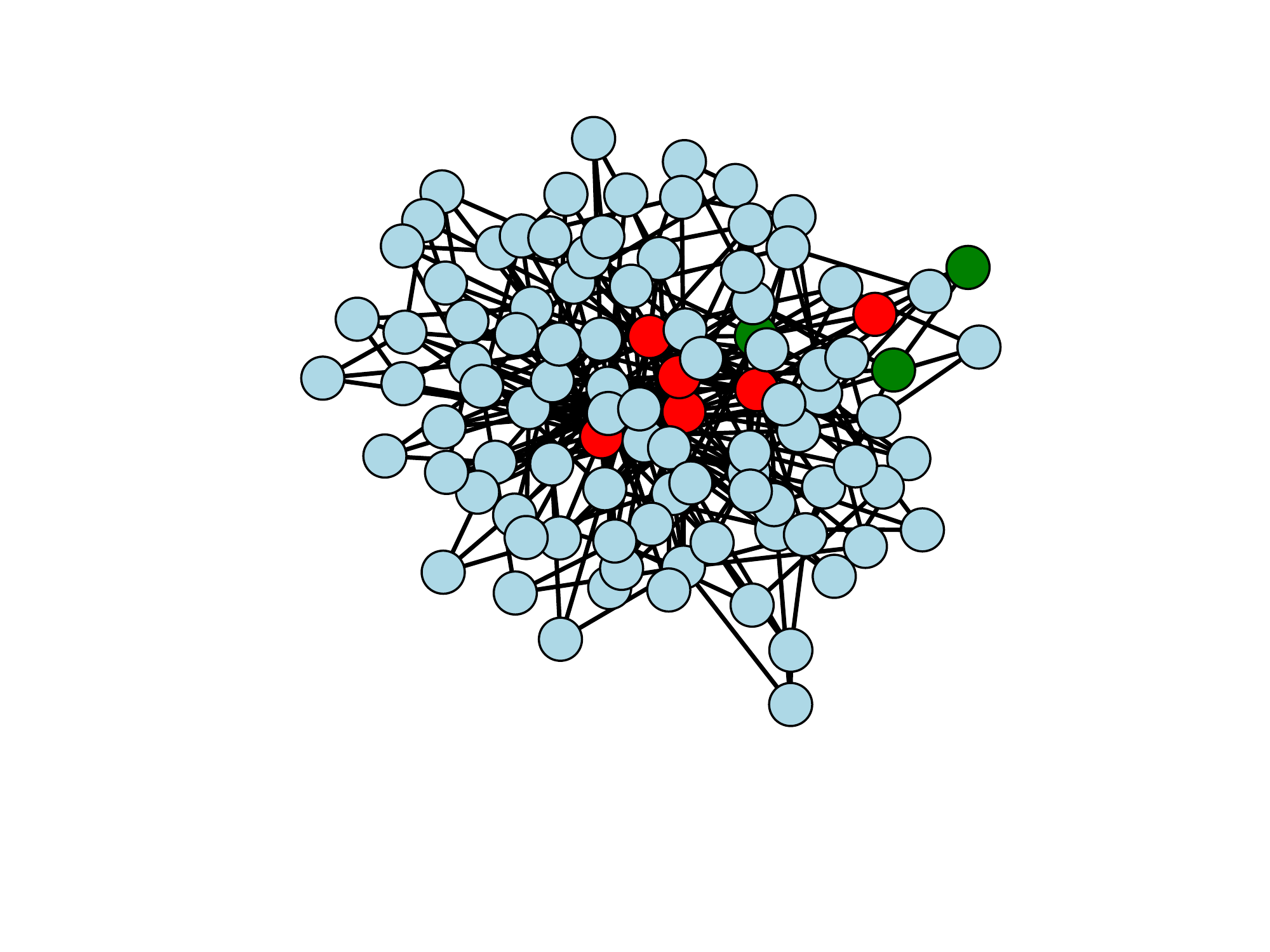}
        \label{1}
    }
    \subfigure[]{
        \includegraphics[width=40mm]{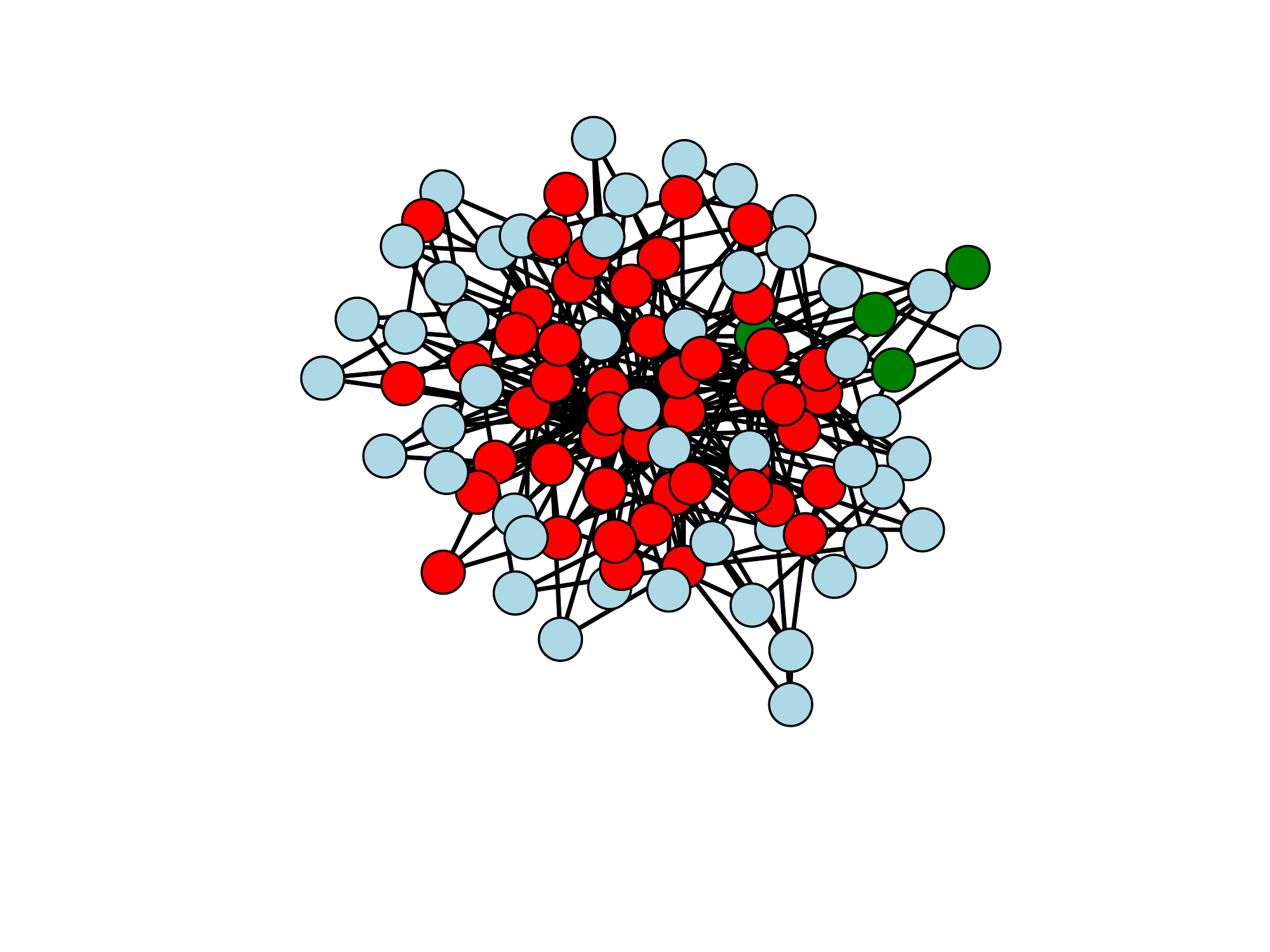}
        \label{2}
    }
    \subfigure[]{
        \includegraphics[width=40mm]{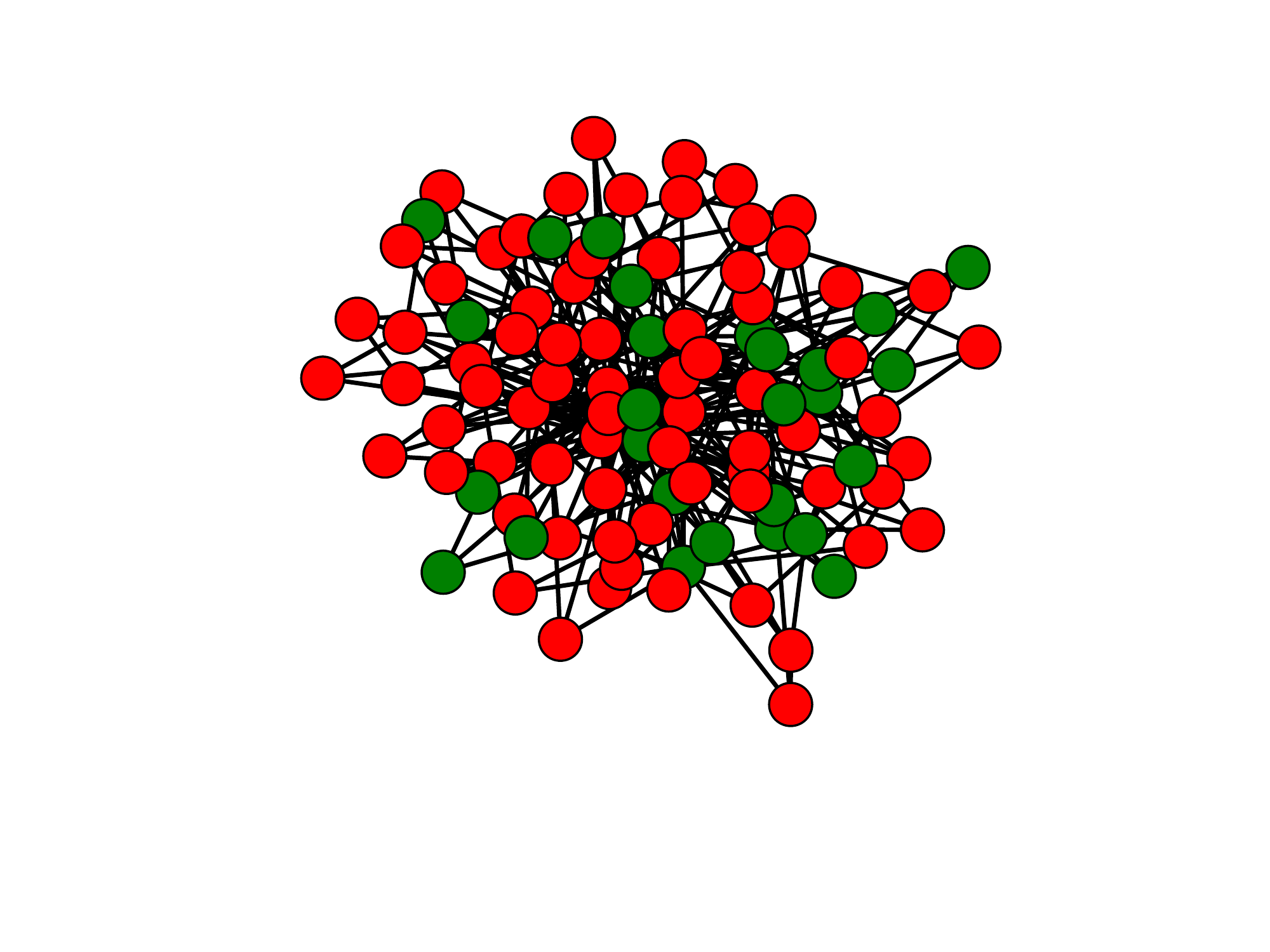}
        \label{3}
    }
    \subfigure[]{
        \includegraphics[width=40mm]{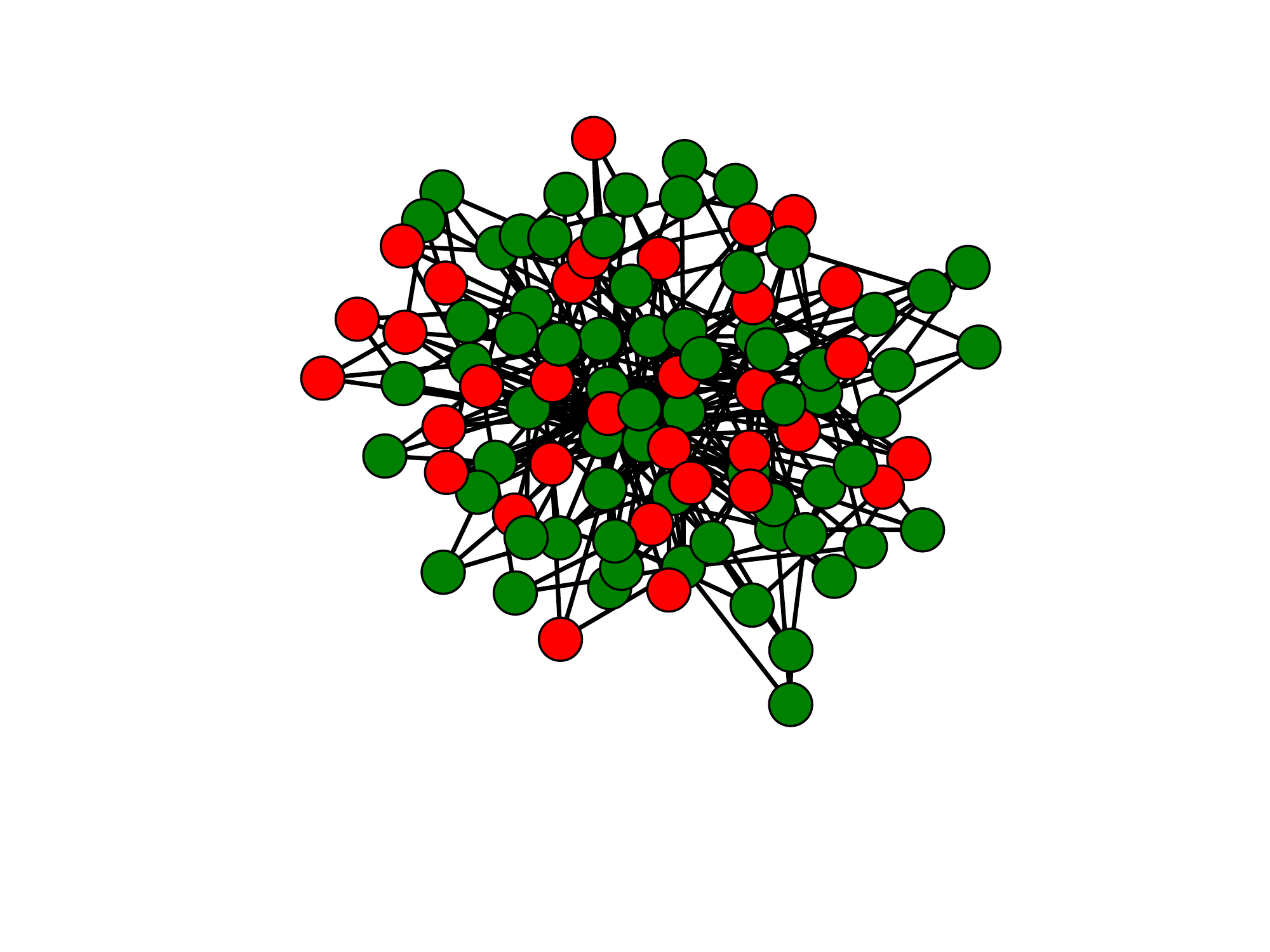}
        \label{4}
    }
\end{center}
\caption{Visualisation of discrete SIR contagion on network, with recovery; Susceptible nodes shown in blue, Infected nodes in red, and Recovered nodes in green.}
\label{SIRvisualisation}
\vspace{-16px}
\end{figure*}
Salath\'{e} and Jones \cite{salathe2010dynamics}, investigating epidemics in networks with community structure, state:
``\emph{An important caveat to mention is that community structure in the sense used throughout this paper (i.e. measured as modularity Q) does not take into account explicitly the extent to which communities overlap. 
\dots the exact effect of community overlap on infectious disease dynamics remains to be investigated}''.
Our first objective is to address this, by studying such `simple' contagions on overlapping community models.

In addition, we will study `complex contagions', as formalized by Centola et al. \cite{centola2007complex,Centola2007a} -- diffusion processes which \textit{require} multiple exposures to a contagion, in order for the contagion to spread.
They argue that complex contagion is a better model for many social phenomena, rather than the single exposure `simple' contagion of the SIR model, and provide a wide survey of sociological literature to back this assertion.
The argument is that for a `high-risk' contagion to diffuse through a social network -- such as the purchase of an expensive product, or participation in a risky political or revolutionary movement -- multiple `social proofs' are needed.
More recently, Romero et al. \cite{romero2011differences} show evidence of complex contagion in Twitter, in the context of hashtag adoption -- this recent empirical work further motivates the study of the speed and scope of complex contagions. %
In addition, there has been some popular debate recently as to the suitability of modern online social networks, for quickly spreading such complex contagions.
The argument is that while these networks have a low average-shortest-path-length (ASPL), this is as a result of random weak ties shrinking the network.
As such, it is argued that any complex diffusion event on such a social network will not spread as fast as a simple one.
Centola writes ``\emph{Our theoretical results also provide new insight into the widely observed tendency for social movements to spread over spatial networks}'' \cite{centola2007complex}, implicitly arguing, after Granovetter \cite{granovetter1973strength} that the social world is made `small' through weak ties, which shrink the underlying spatially clustered social network, but which do not carry a complex contagion quickly.

Indeed, in a model of the world where non-overlapping communities are only weakly connected to each other, by random ties -- or even in a Watts-Strogatz model of the world, where the network is heavily clustered, but the low ASPL depends on randomly re-wired links -- this argument makes intuitive sense.
However, we argue that the overlapping nature of community in online social networks means contagions can still spread fast through community structure.
We argue that short paths through overlapping communities exist on such networks, allowing complex and simple contagions to spread fast, simply through community overlap.
This is an important topic of study, as a necessary prerequisite to developing algorithms for a range of important problems, such as which links are most important to the spread of diffusion -- such as new product adoption on a network -- or which groups of influential users should be mined.
We present results that investigate the effect of overlapping community structure on simple contagions; we then investigate the effects of community structure -- both overlapping and non-overlapping -- on the spread of complex contagions; and finally we conduct structural analysis and simulation on empirical networks.

\subsection{Other relevant literature}
Other researchers \cite{shi2007networks,van2006strong}, have investigated a limited number of social network datasets, and found evidence that paths through these networks exist, traversing only `strong ties', consistent with a model of overlapping community.
However, the spread of contagions on such networks has not -- to our knowledge -- previously been examined in detail.
Communities on many real social networks have been found to overlap pervasively.
Leskovec et al. \cite{leskovec2008statistical} have shown that \textit{distinct} communities are hard to find in the cores of larger networks, at larger scales, again consistent with a world view in which communities overlap pervasively.
Reid et al. \cite{reid2011partitioning} have shown that on some social networks, cliques and communities exist in the boundaries between other communities in a widespread manner, consistent with a model of networks containing pervasively overlapping communities.
There is also a growing body of literature \cite{ahn2010link,evans2009line} dedicated to finding specific types of overlapping community in empirical networks; we constrain our analysis to synthetic models of overlapping network, and to simulation of contagions on real data which is known to have overlapping structure; we do not examine specific overlapping community finding algorithms to extract specific types of community.

\section{Modelling}

\subsection{Contagion Models}
The SIR model is a long standing mathematical model of epidemic spread \cite{anderson1979population, kermack1150}.
In this model, individuals are divided into three `compartments' -- those who are Susceptible to, Infected by, and Recovered from the contagion.
This simple mathematical model assumes that the population mixes homogeneously, and that the chance of an individual becoming infected is proportional to the contact and infection rate of the disease, and the number of `Infecteds' currently in the population.
The assumption of homogeneous mixing underlying this model is, of course, a profound simplification.
Recent modelling approaches have investigated the behavior of epidemics by modelling the contact network of each individual \cite{pastor2001epidemic,Newman2002a,Meloni2009}.
In such approaches, the study of the contagion in the population becomes the study of the contagion across the contact network.
Such models can have rich dynamics and epidemic thresholds different from a simple homogeneous mixing model; the topological structure of networks plays an important role in the spread of contagions on them.

One common way of studying the behavior of SIR-like contagions on networks is to stochastically simulate the contagion as a discrete process across the network.
In such simulations, the network is typically initialized to a state where all nodes are susceptible.
A randomly chosen starting node is then infected, and the simulation is started.
At each simulation time period, each infected node infects each susceptible neighbor according to a probabilistic model.
The chance of a susceptible node becoming infected is thus a function of the number of infected nodes $i$ in contact with it, and the contagion parameter $\beta$.
In Liu et al. \cite{liu2005epidemic} and others, this contagion parameter is used to directly calculate the probability of infection, yielding the infection function $1-(1-\beta)^{i}$.  
Salath\'{e} and Jones \cite{salathe2010dynamics} use the function $1-\exp(-\beta i)$.
Both functions yield very similar infection curves.
For simplicity we will use the same function as Salath\'{e} and Jones, some of whose work we replicate.
Infected nodes recover from infection, with probability $\gamma$ at each time period, after which they become `Recovered' and not able to infect any more surrounding nodes. It is important to note that while exposure to multiple sources of infection ($i$) increases the probability of being infected, in the SIR model it is possible to become infected so long as at least one exposure has occurred ($i\ge1$).
Such simulations are typically repeated many times, and the speed and size of the contagions quantified and studied.
Figure \ref{SIRvisualisation} shows a visualization of the evolution of an epidemic from such a model, as simulated on a small Barabasi-Albert network.
It is this modelling approach we use in our experiments, to investigate simple SIR contagions.

Centola et al. \cite{centola2007complex} explore the idea of a `complex' contagion.
In such a contagion, exposure from more than a single source is necessary for the contagion to spread. That is, for infection to occur, it is necessary that $i \ge \alpha$, where $\alpha>1$ is the \emph{threshold of activation}.
They do not model a `recovered' state, nor do they model a probabilistic chance of contagion once the threshold of `activated' or `infected' neighbors is reached; rather, once this threshold is reached, the contagion spreads.
We use this same modelling approach in our experiments.
We focus specifically on modelling complex contagions where the threshold of activation, $\alpha$, is 2.
This is the simplest possible complex contagion, and means that a node requires two neighbors that have already been infected before that node becomes infected.
We examine this threshold value, to investigate the possibility of \textit{any} complex contagion propagating fast, given that the standard theory of social networks would expect such contagions to spread slowly, as they cannot cross the `weak-tie' `shortcuts' which make the network small.
Due to the nature of complex contagions, in order to start the contagion in our simulations, we must infect more than a single starting node. After Centola et al. \cite{centola2007complex}, we randomly select a single `focal' node, and then initialize that node and its neighbors to the infected state, to start the contagion process.

\subsection{Network Models}
We now investigate simple and complex contagion on several models which feature overlapping, and non-overlapping, community structure.
To begin, we replicate the network model of Salath\'{e} and Jones \cite{salathe2010dynamics} -- henceforth referred to as the `$SJ$ model'.   
This model makes 50 distinct, disjoint, non-overlapping communities. Each community is a Watts-Strogatz ring lattice of 40 nodes and 160 edges, with each node connected to 8 neighbors.  This  yields a network of 2,000 nodes in total and 8,000 \textit{intra}-community edges.  Next, 2,000  \textit{inter}-community edges are added at random to the network, thus joining the communities by weak ties. Finally, to yield a set of  community structures of varying modularity, some proportion of the inter-community edges are re-wired into the communities to become intra-community edges and thus increase the modularity of the network.  
We might question how realistic a model  of community structure this is: communities are generally thought to be quasi-clique, rather than lattice-like. 
However, the fact that communities are internally dense is captured, which is the most important attribute of modelling community structure. %

We aim to study diffusion in a network model that is similar, but has community overlap. 
We want to ensure that our results are as comparable as possible to those obtained on the non-overlapping $SJ$ model, and thus seek an overlapping model with the same number of nodes and edges as the $SJ$ model, to allow direct comparison of results.
To proceed, we create communities using the same Watts-Strogatz ring lattice of 40 nodes and 8 neighbors each and following the $SJ$ model, start with 50 such communities. To introduce overlap into these 50 communities is not possible without either increasing the number of community edges, or decreasing the number of nodes covered. 
Our solution is to introduce community overlap to the model in the re-wiring step.

Specifically, starting with 2,000 random inter-community edges, as in the $SJ$ model, some proportion of these edges is rewired -- but rather than re-wiring into the existing communities, instead they are re-wired to form new overlapping communities: for each set of 160 re-wired edges, 40 nodes are chosen at random and the edges are created to form a new Watts-Strogatz ring lattice of degree 8 between these nodes.  %
Thus where in the $SJ$ model, each 160 inter-community edges are rewired to be intra-community edges, we instead create an extra ring lattice community that overlaps with the existing communities.
Now we have a parameterized model, very similar to the $SJ$ model -- in that it contains the same number of nodes and edges -- but with which we can control the extent to which the edges exist as random between-community edges, or as overlapping community edges.

This is not a perfect model: communities are still ring-lattice like, as each overlapping community is identical to the ring lattice communities in the original model.
We would also not necessarily argue that the existence of overlapping communities should reduce the amount of random inter-community edges.
However, this simple change to the model provides us with a benchmark similar to the original, thus allowing our results to be compared with it; while also featuring a parameterized amount of community overlap, allowing us to investigate the effect of this overlap.
We refer to this modified model as the $SJ_{o}$ model.\footnote{Datasets at: sites.google.com/site/diffusionnetworkoverlap}

\section{Experiments}

\subsection{SIR contagion}
After Salath\'{e} and Jones \cite{salathe2010dynamics} we now study the effect of community structure, on the speed and reach of simulated epidemics.
Where they studied solely non-overlapping community, we will also investigate our simple modification of their model which introduces overlap.
It is necessary to perform these experiments in detail, before drawing conclusions about diffusion behavior on these networks, due to the unpredictable nature of epidemic behavior.

They investigate the relationship of community structure, with the final size, duration, and peak prevalence of a contagion.
They do this by calculating the modularity value $Q$ of the network, as found by a spin glass optimisation method, and relate it with the diffusion characteristics.
However, many concerns have been voiced over the suitability of the modularity measure for quantifying community structure, and it is inappropriate for investigating overlapping community structure. As Good et al. write ``\emph{These results imply that the output of any modularity maximization procedure should be interpreted cautiously in scientific contexts}'' \cite{good2010performance}.
We thus investigate the synthetic networks, not according to the modularity found on them by a particular algorithm, but instead by the number of edges rewired in their construction.
In any case, in this particular benchmark, the modularity $Q$ of the network in the \textit{original} $SJ$ model corresponds directly to the proportion of edges rewired.

After \cite{salathe2010dynamics}, there are several key attributes of the epidemic we are interested in analyzing.
One is the \textit{final size} of the epidemic -- this is the total number of recovered and infected nodes; the total number who the epidemic infected at any time.
We are also interested in the \textit{duration} -- how long did it last for?  Salath\'{e} and Jones define this as the time until the last node is recovered.
However, this is highly sensitive to the recovery probability $\gamma$, and is not defined when we do not have a recovered state; so when considering complex contagions, we will define the duration to be the time at which the last node was infected.
Another property studied is that of the \textit{peak prevalence} -- what was the peak number of individuals in the infected state at any one time?  In a situation where nodes do not recover, this is identical to the final size.
Finally, we consider the \textit{`speed'} of the contagion to be a feature of interest, and we define this as the length of time before the the epidemic has reached a size (not prevalence) of at least 50\% of the population.

\begin{figure*}[p]
    \subfigure[]{
        \includegraphics[width=90mm]{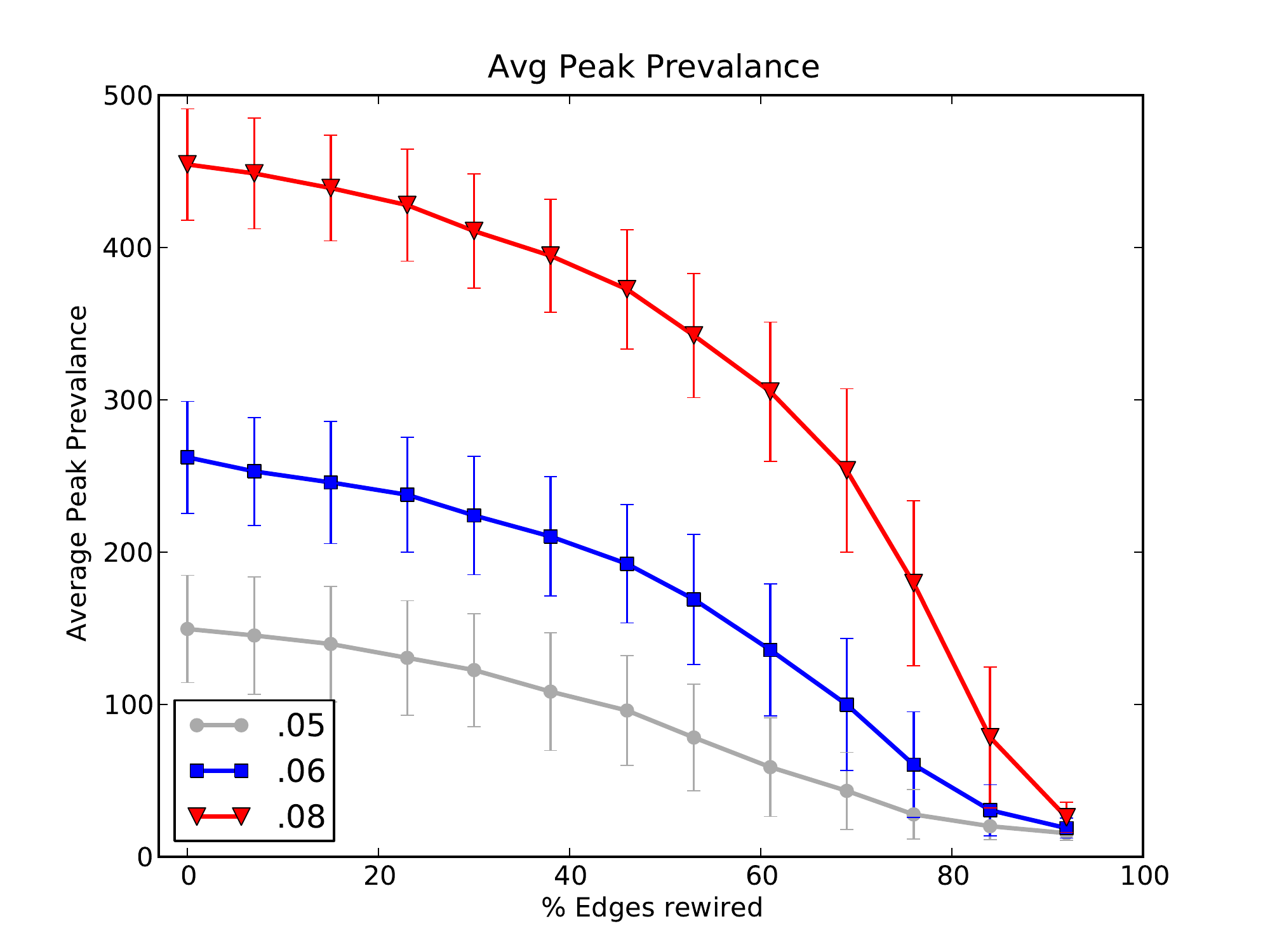}
        \label{AvgPeakPrevalanceSJ}
    }
    \subfigure[]{
        \includegraphics[width=90mm]{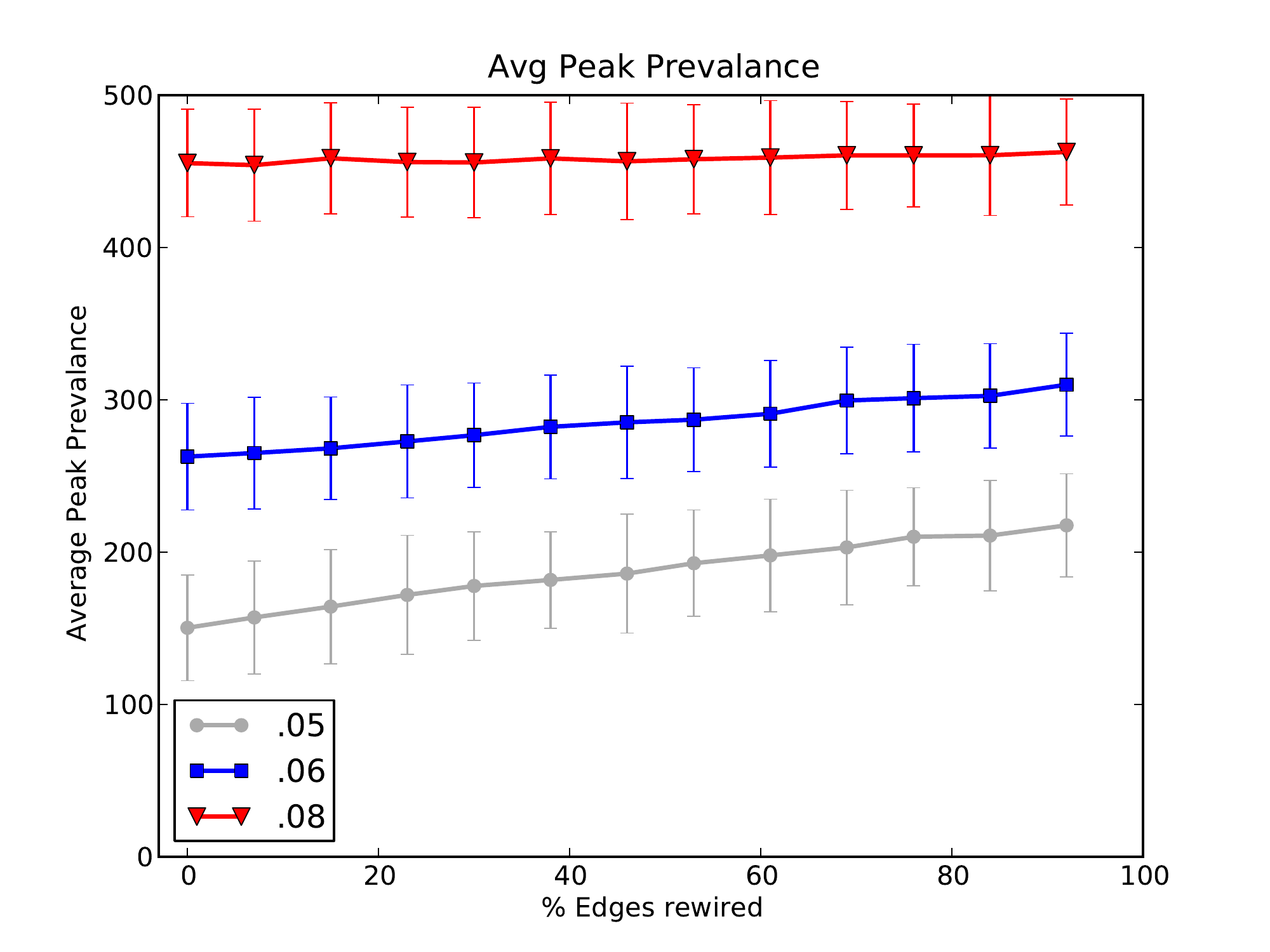}
        \label{AvgPeakPrevalanceOverlapSimple}
    }
\caption{Average peak prevalence for simple contagion, replicating results of \cite{salathe2010dynamics} (a) and presenting \newline corresponding results as community overlap increases (b).}
    \subfigure[]{
        \includegraphics[width=90mm]{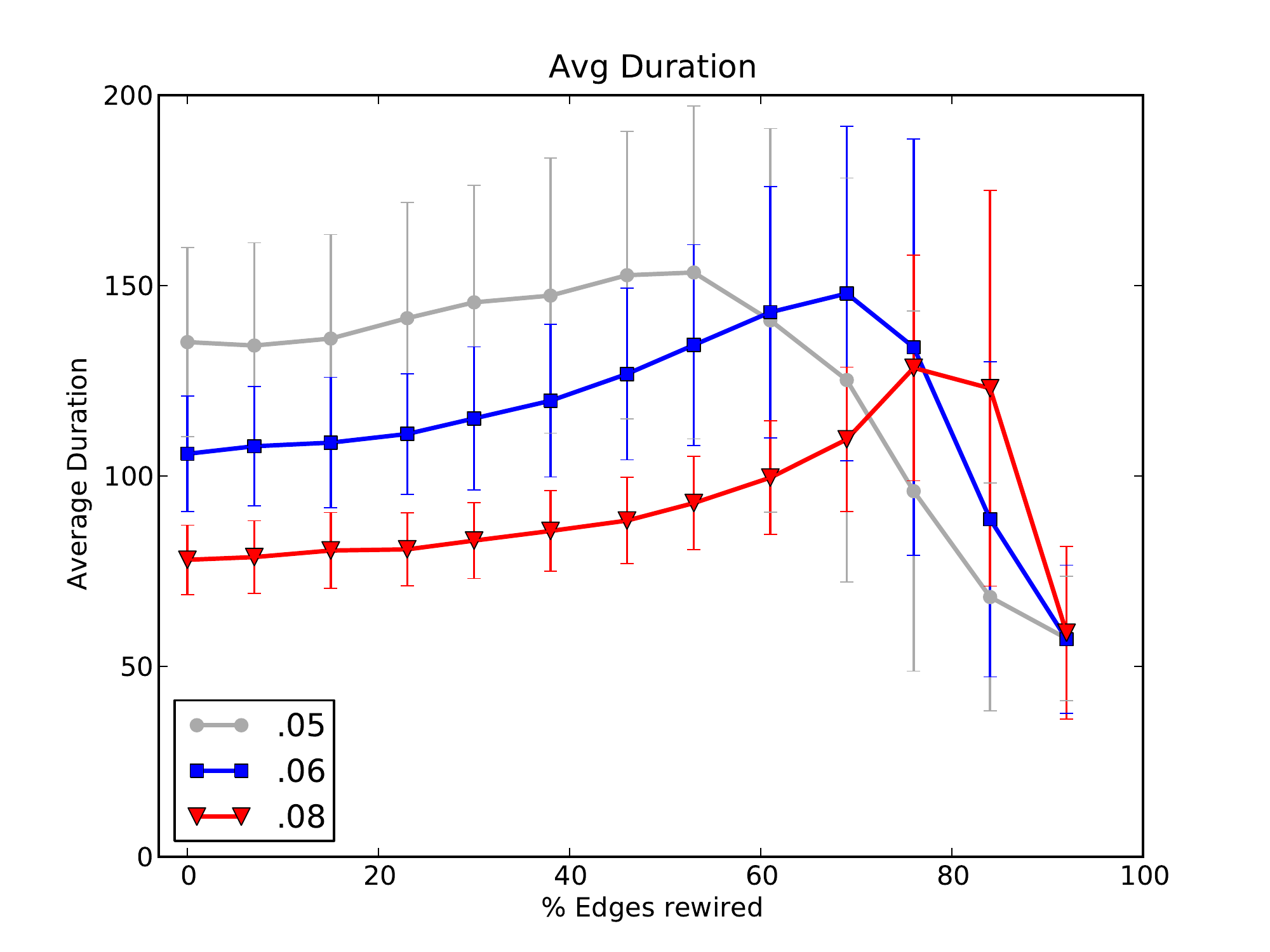}
        \label{AvgDurationSJ}
    }
    \subfigure[]{
        \includegraphics[width=90mm]{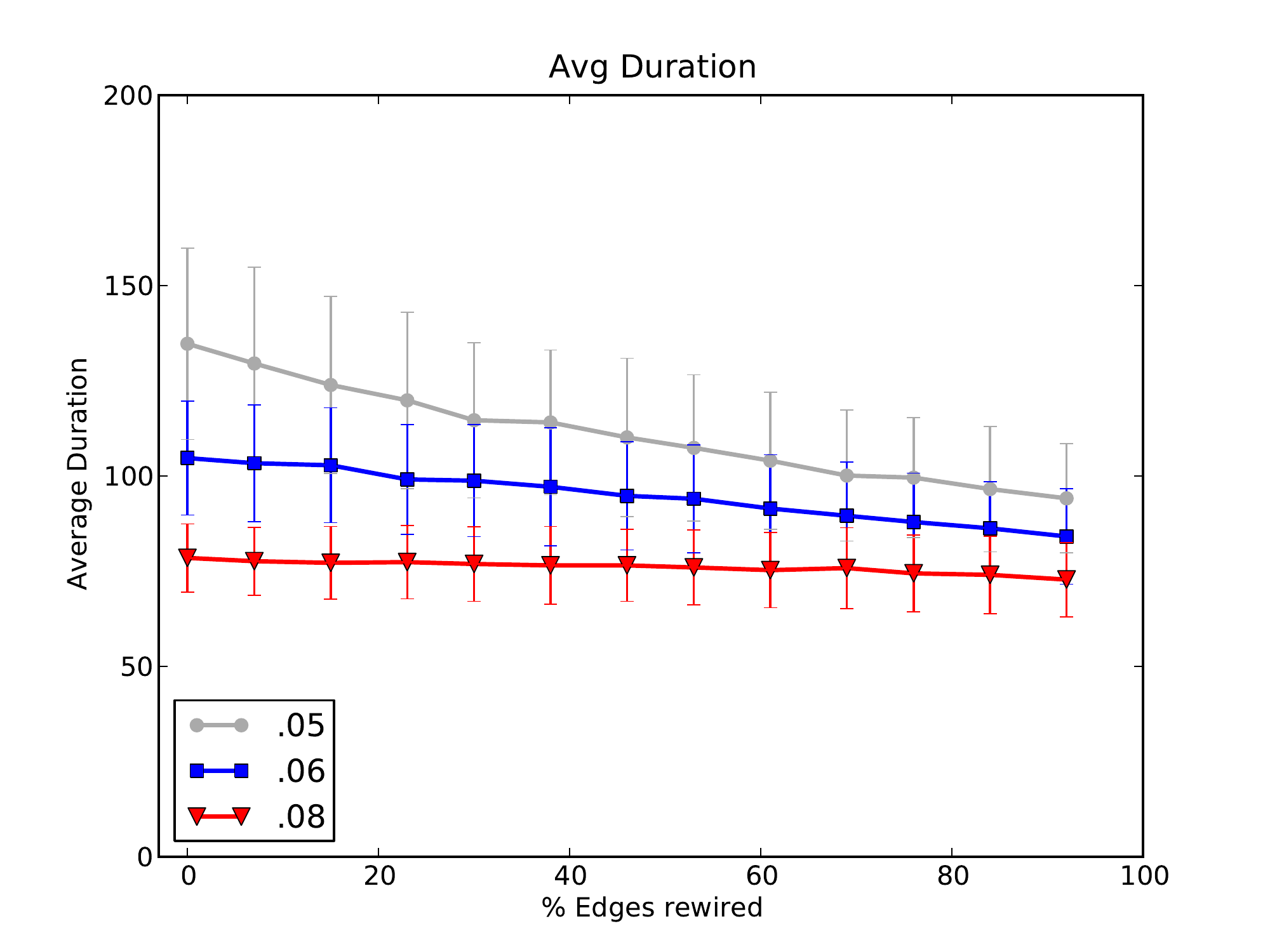}
        \label{AvgDurationOverlapSimple}
    }
\caption{Average duration for simple contagion, replicating results of \cite{salathe2010dynamics} (a) and presenting corresponding results as community overlap increases (b).}
    \subfigure[]{
        \includegraphics[width=90mm]{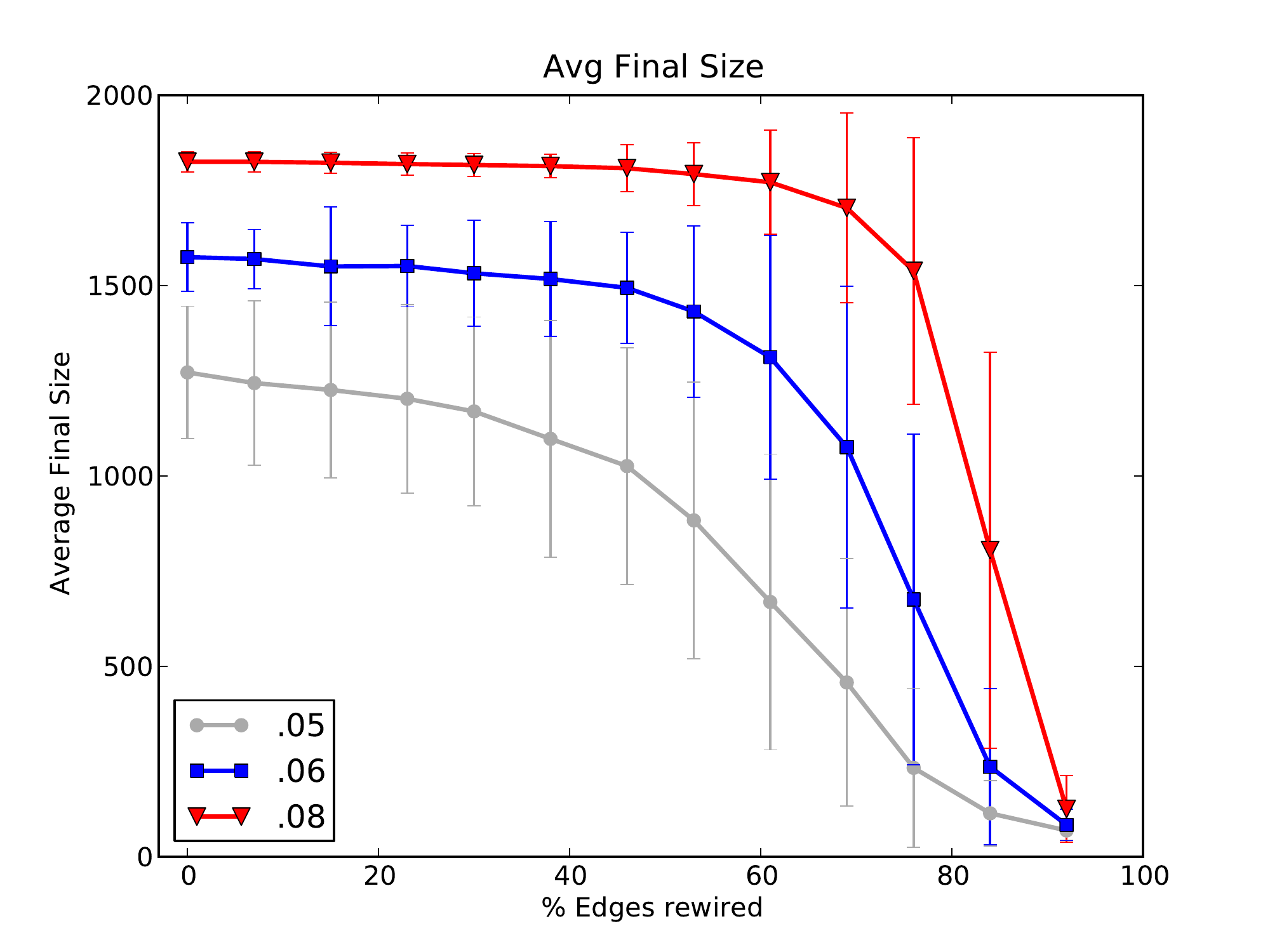}
        \label{AvgFinalSizeSJ}
    }
    \subfigure[]{
        \includegraphics[width=90mm]{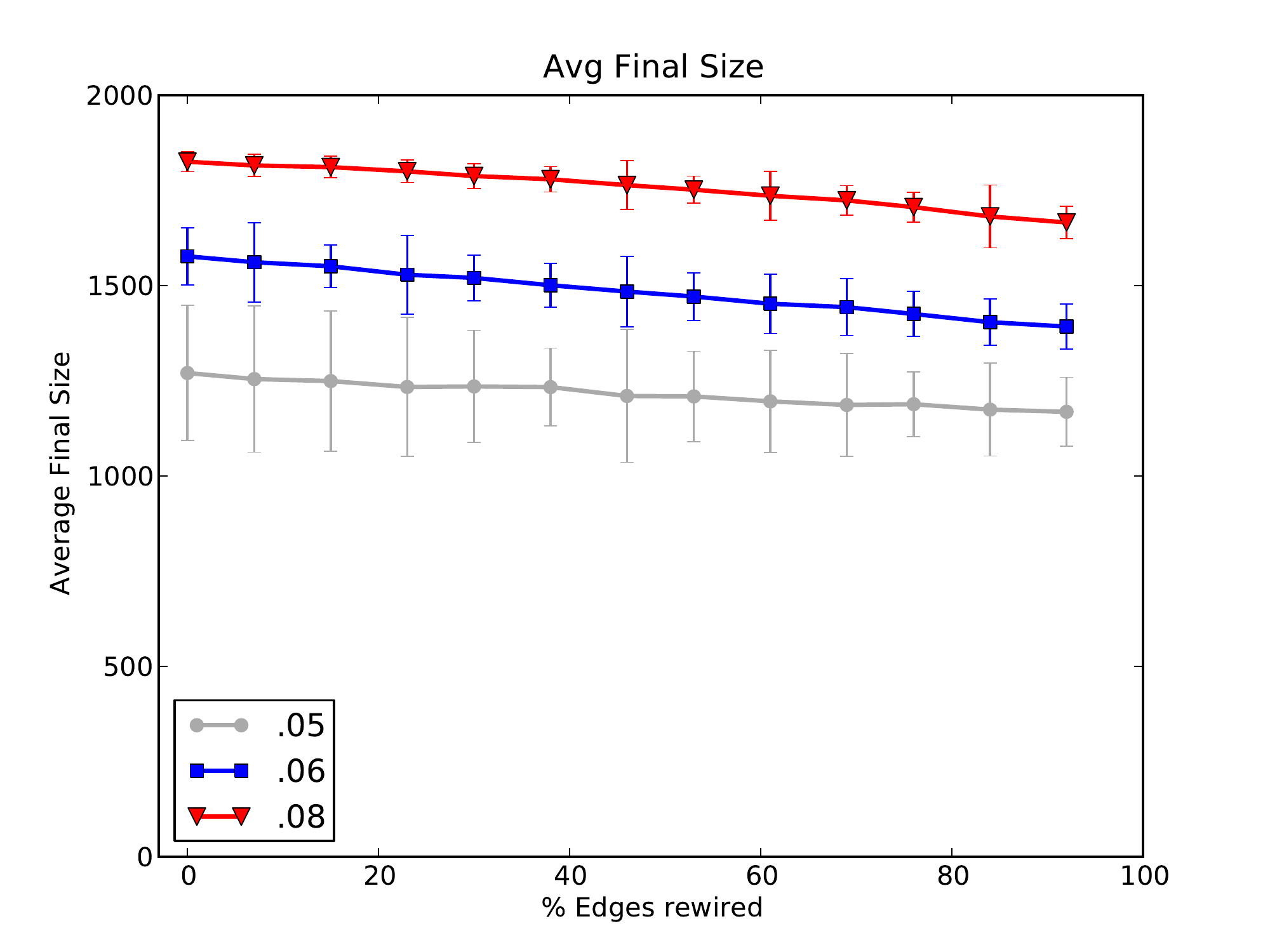}
        \label{AvgFinalSizeOverlapSimple}
    }
\caption{Average final size for simple contagion, replicating results of \cite{salathe2010dynamics} (a) and presenting corresponding results as community overlap increases (b).}
\end{figure*}

Our first experimental step in this investigation is to replicate the results of Salath\'{e} and Jones, as they investigate the effect of the proportion of inter vs intra community edges on their benchmark.
In all our SIR simple contagion modelling, we use their values of $\beta$ (0.05, 0.06, 0.08) and $\gamma$ (0.2).
Further, for comparability, as in their work, we discard all results where less than 5\% of nodes were infected by the contagion; only contagions which have got a foothold are of interest.
We first briefly present the results of our replication, in Figures \ref{AvgPeakPrevalanceSJ} \ref{AvgDurationSJ} and \ref{AvgFinalSizeSJ}.  The same results can be seen here: with non-overlapping communities, the increased proportion of intra-community edges reduce the size, prevalence, and speed with which the epidemic spreads.

However, on our $SJ_{o}$ model, with which we investigate the effect of overlap on the same contagions, we find very different results, as shown in Figures \ref{AvgPeakPrevalanceOverlapSimple} \ref{AvgDurationOverlapSimple} \ref{AvgFinalSizeOverlapSimple}.
As we re-wire edges into additional ring lattice communities, identical to the existing communities, except that their nodes overlap, we do not find the same decrease in epidemic spread that occurs with increased non-overlapping community structure.
These results show that unlike non-overlapping community structure, overlapping community structure does not act as an effective `brake' to the spread of contagions.
Rather, the final size of contagions examined remains largely static as random edges are re-wired into overlapping communities, with the increased overlapping community structure showing but a slight decrease in epidemic duration, and a slight increase in prevalence.
These results show that while non-overlapping community structure may slow the spread of contagion, overlapping community structure does not necessarily have the same effect.

We note here that the average values shown in our graphs have substantial error bars; these represent the standard deviation, not the standard error.
It is the nature of epidemic modelling that there will be substantial deviation in the properties of simulated epidemics; each individual point, is, however the result of 2,000 simulations, each on a different realization of the network, which means the standard error of the mean is too small to plot; the variance is large, but the trends of the mean are robust.

\subsection{Complex contagion on community models}
\begin{figure*}[t]
    \subfigure[]{
        \includegraphics[width=90mm]{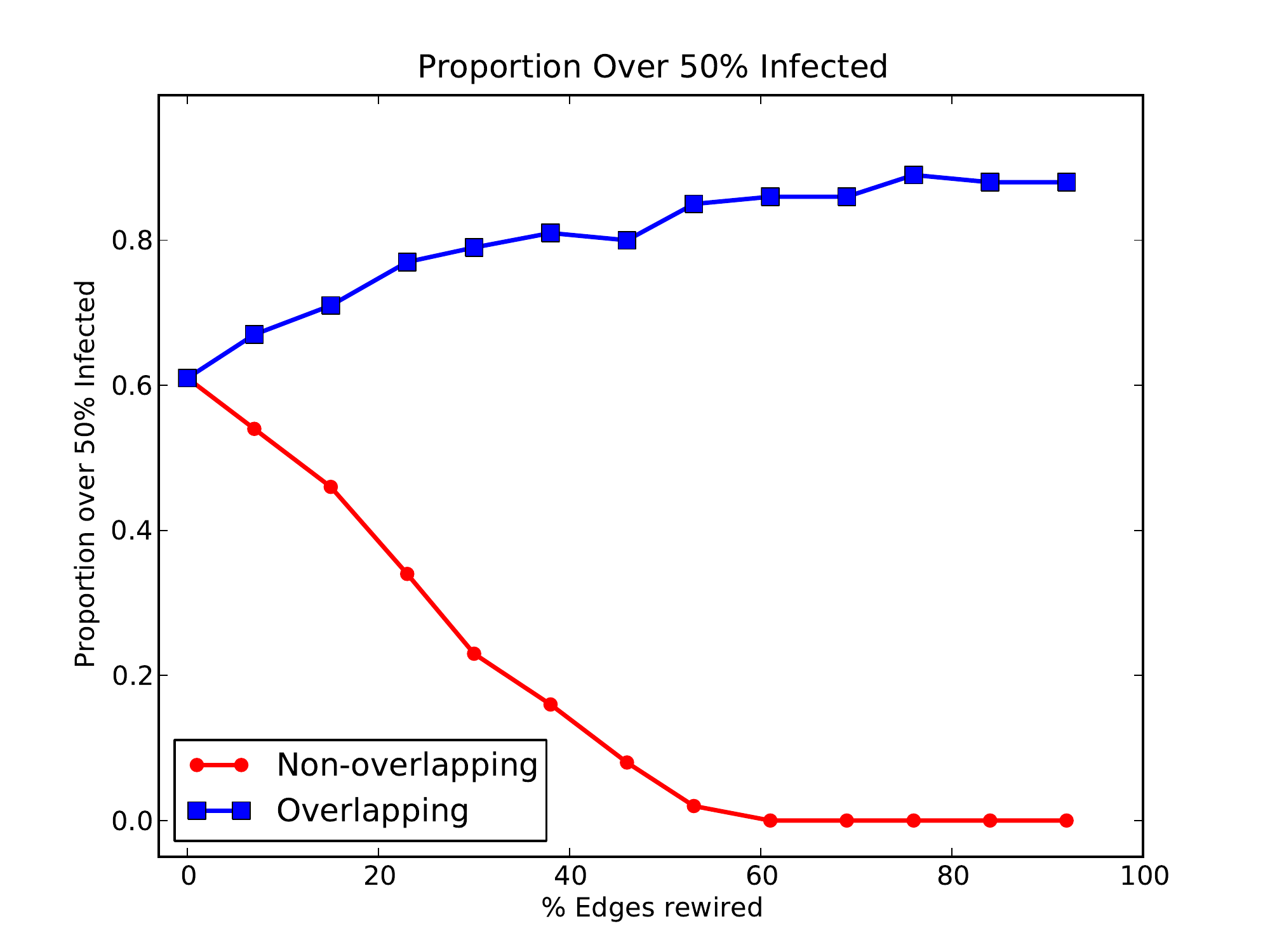}
        \label{ProportionInfectionsOver50pc}
    }
    \subfigure[]{
        \includegraphics[width=90mm]{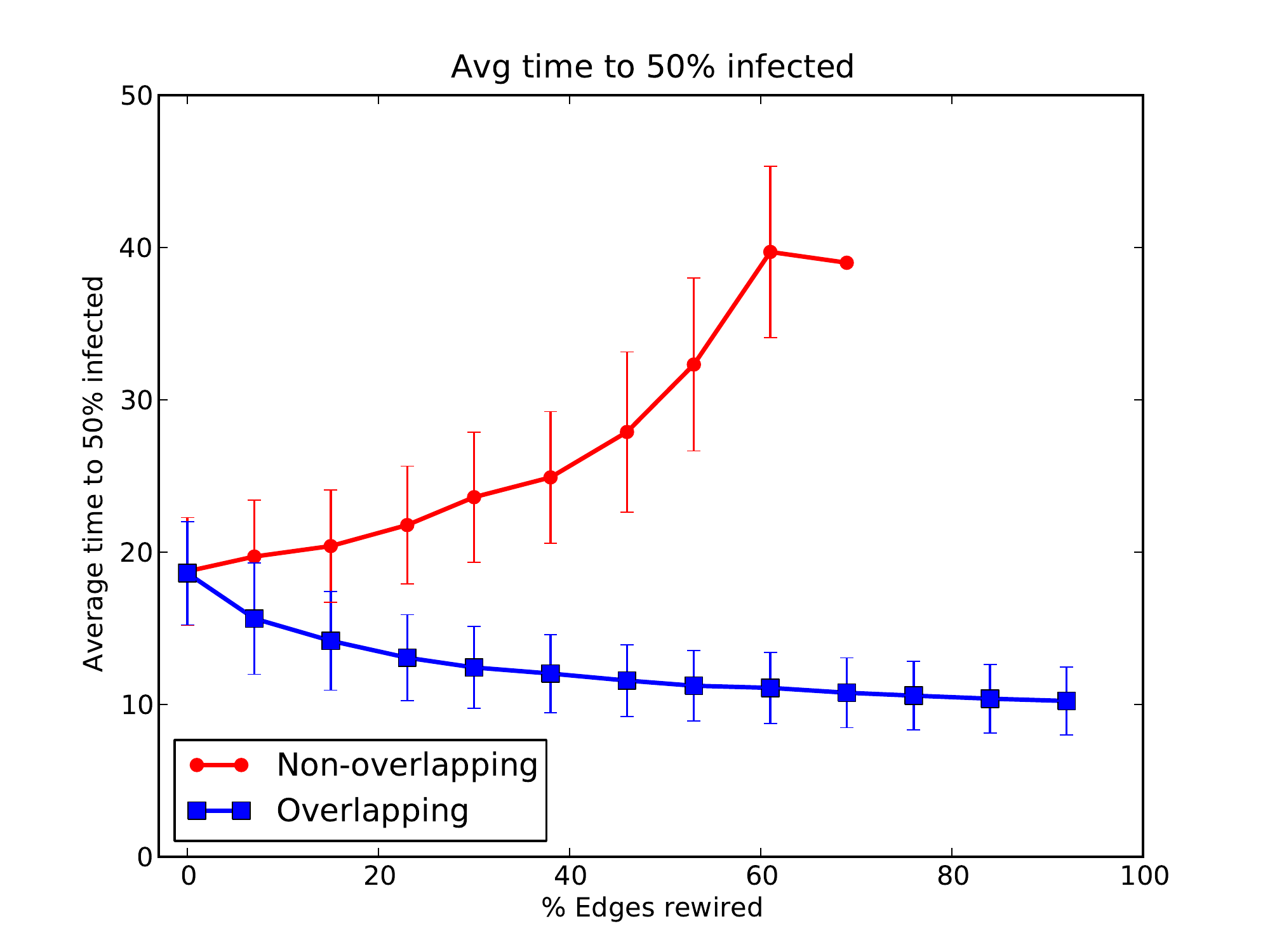}
        \label{Avgtimeto50pcinfectedAvgComplexOverlapVsNonOverlap}
    }
\caption{Complex contagion on overlapping and non overlapping models, as edges are rewired. Figure (a) shows the proportion of contagions that infected over 50\% of nodes. Figure (b) shows, for these contagions, the average time until 50\% of the nodes are infected. (There is no Recovery in our complex contagion model; hence the primary attributes are whether the contagion can become established, and how fast it spreads.)  Increasing non-overlapping community structure slows the complex contagion propagation; however, increasing overlapping community structure actually speeds it up.}
\label{complexContagionSJ}
\vspace{-12px}
\end{figure*}
\begin{figure}[!htb]
    \includegraphics[width=90mm]{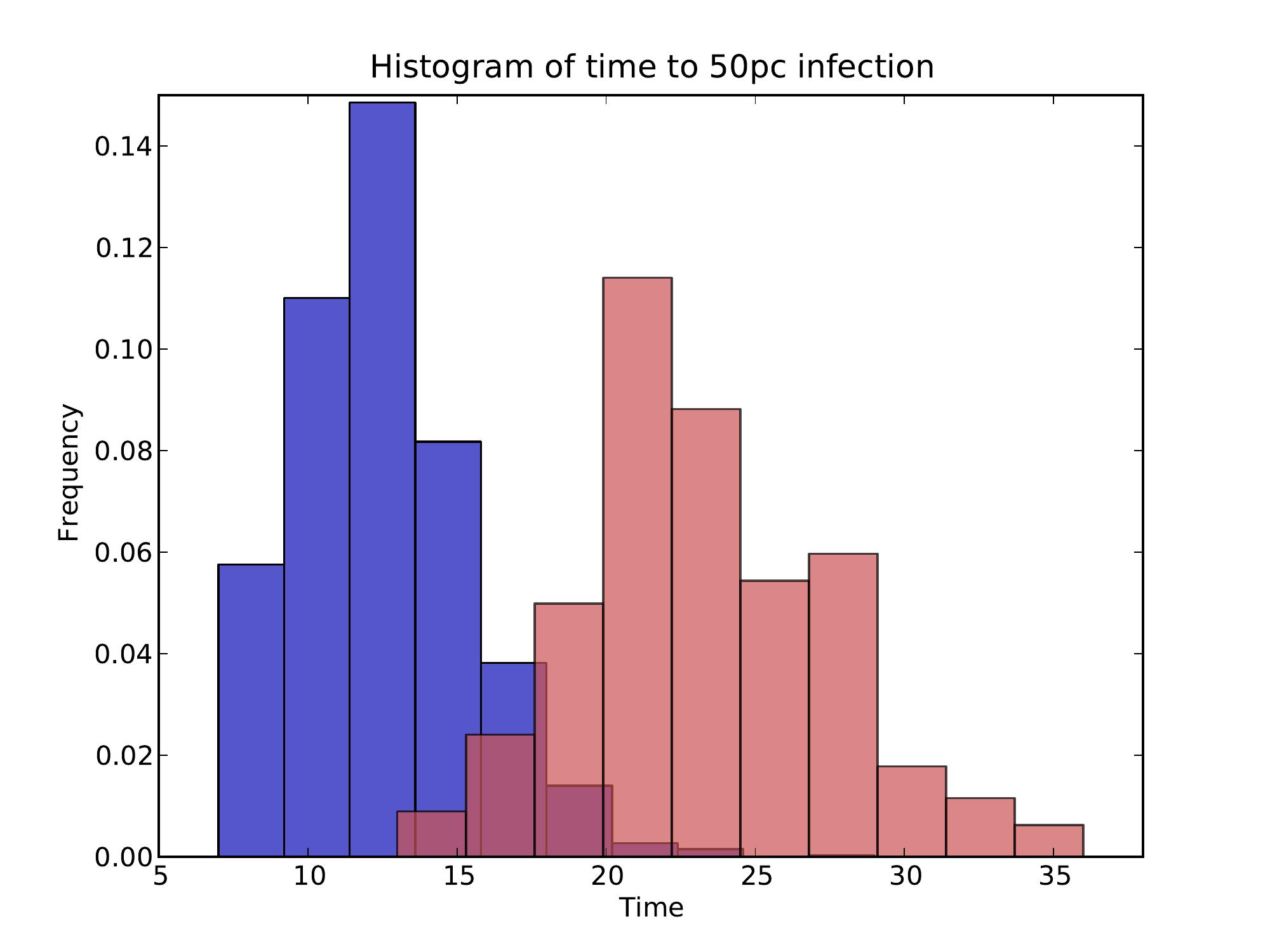}
\caption{Histogram of time to 50\% infected for complex contagion on overlapping (blue) and non-overlapping (light red) models, at 32\% rewiring. Data from 2,000 simulations on each model.}
\label{histogramOfTimeTo50pcInfectComplex}
\vspace{-18px}
\end{figure}

We now conduct an experiment similar to that of modelling simple contagion on overlapping structure, but instead use a complex contagion which requires exposure from two different neighbours for a node to become infected.
We conduct our experiment on both the original $SJ$ model and the $SJ_o$ model, and present our results in Figure \ref{complexContagionSJ}.
As there is no recovery in our complex contagion model, the peak prevalence will always be the final size, and the lack of recovery means the final size will be of less interest than the speed of the propagation.
Hence we focus on whether the network topology will allow the complex contagion to spread widely, and, if it does, the speed with which the contagion infects nodes.
As shown in Figure \ref{ProportionInfectionsOver50pc} the proportion of realisations in which complex contagion infects over 50\% of the nodes falls as the non-overlapping community structure is increased; however, it rises as the overlapping community structure is increased.
Further, it can be seen in Figure \ref{Avgtimeto50pcinfectedAvgComplexOverlapVsNonOverlap} that the increasing non-overlapping community structure slows the complex contagions that do propagate; however, perhaps surprisingly, increasing overlapping community structure actually speeds them up in a quantifiable way.
As each point on these charts is an aggregation of 2,000 simulations, to show more clearly the distributions behind the aggregate plot, in Figure \ref{histogramOfTimeTo50pcInfectComplex} we break out the time to 50\% infected, for the 32\% rewiring case.
While the histograms do overlap in this diagram, the different trends in the infection speed for the overlapping and non-overlapping models are clear, with the overlapping model showing much faster spread.
These results show that there is a clear difference between the effect of increased overlapping vs non-overlapping community structure, on this particular model.

This is just one model of overlapping community, where we have made some of the simplest modifications possible to the $SJ$ model that feature community overlap.
But the very different diffusion results on this model illustrate a key point: we know that real community structure overlaps -- given that, and our results here, it is not reasonable to conclude that increased community structure acts as a barrier to the spread of contagion; if the increased community structure is overlapping, it may even increase the speed of contagion, as against other previously examined models.

Having investigated contagion in detail on one particular model, we now consider how general our results are.
Network models with overlapping community structure is still an active area of research, however several have been proposed, including a family of models which view real world networks as projections of bipartite community-node graphs \cite{guillaume2006bipartite} \cite{botha2010community}.
These models typically produce networks of communities which are inherently overlapping; and have explanatory power as intuitive models of real world networks.
They produce networks that are both clustered, and have a low ASPL, not by adding random `weak-ties', but instead by creating overlapping communities in such a way that short paths exist through these overlapping community structures -- like our $SJ_o$ model.

\begin{figure*}[!htb]
\begin{center}
    \subfigure[]{
        \includegraphics[width=42mm]{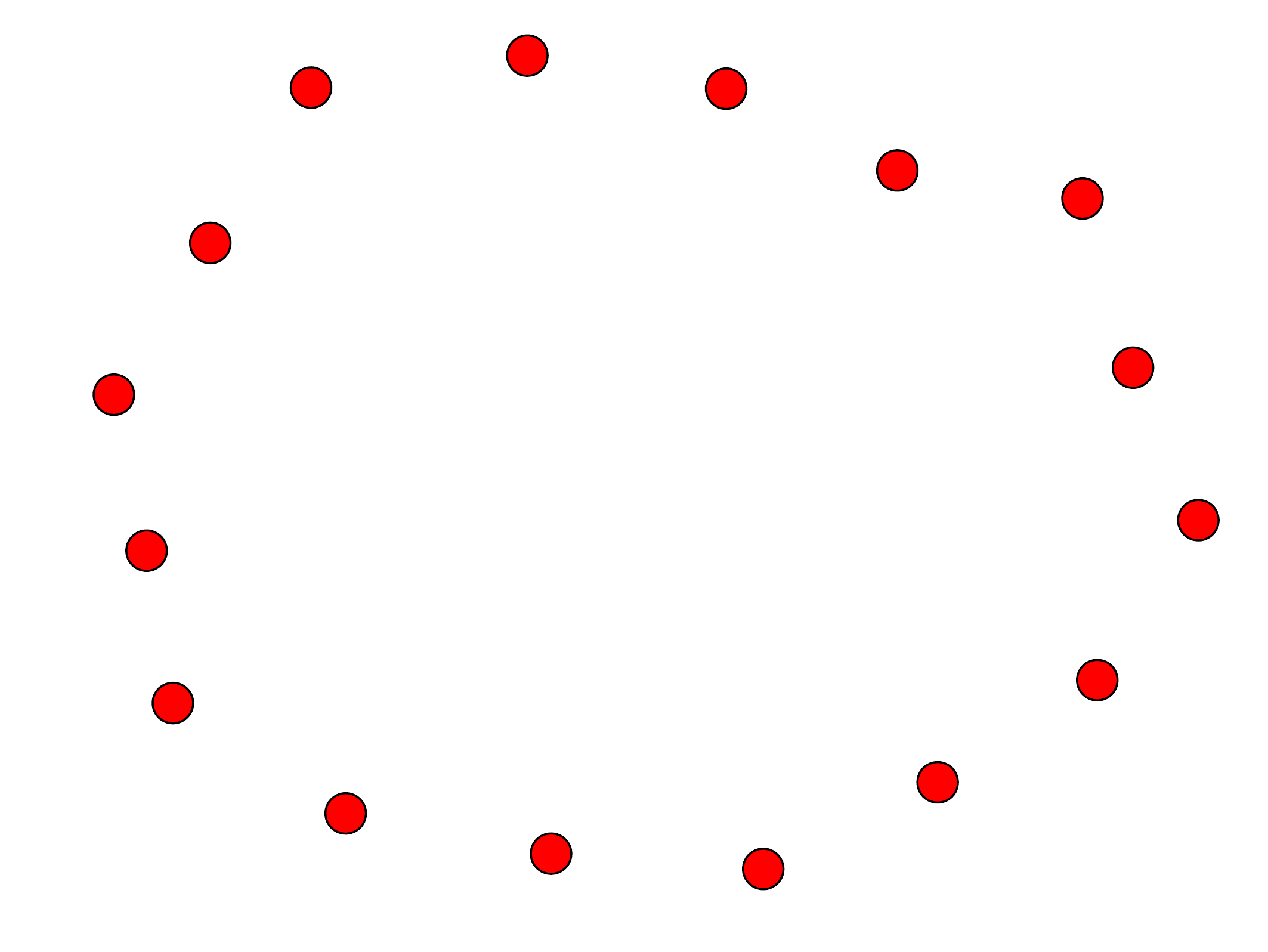}
        \label{1}
    }
    \subfigure[]{
        \includegraphics[width=42mm]{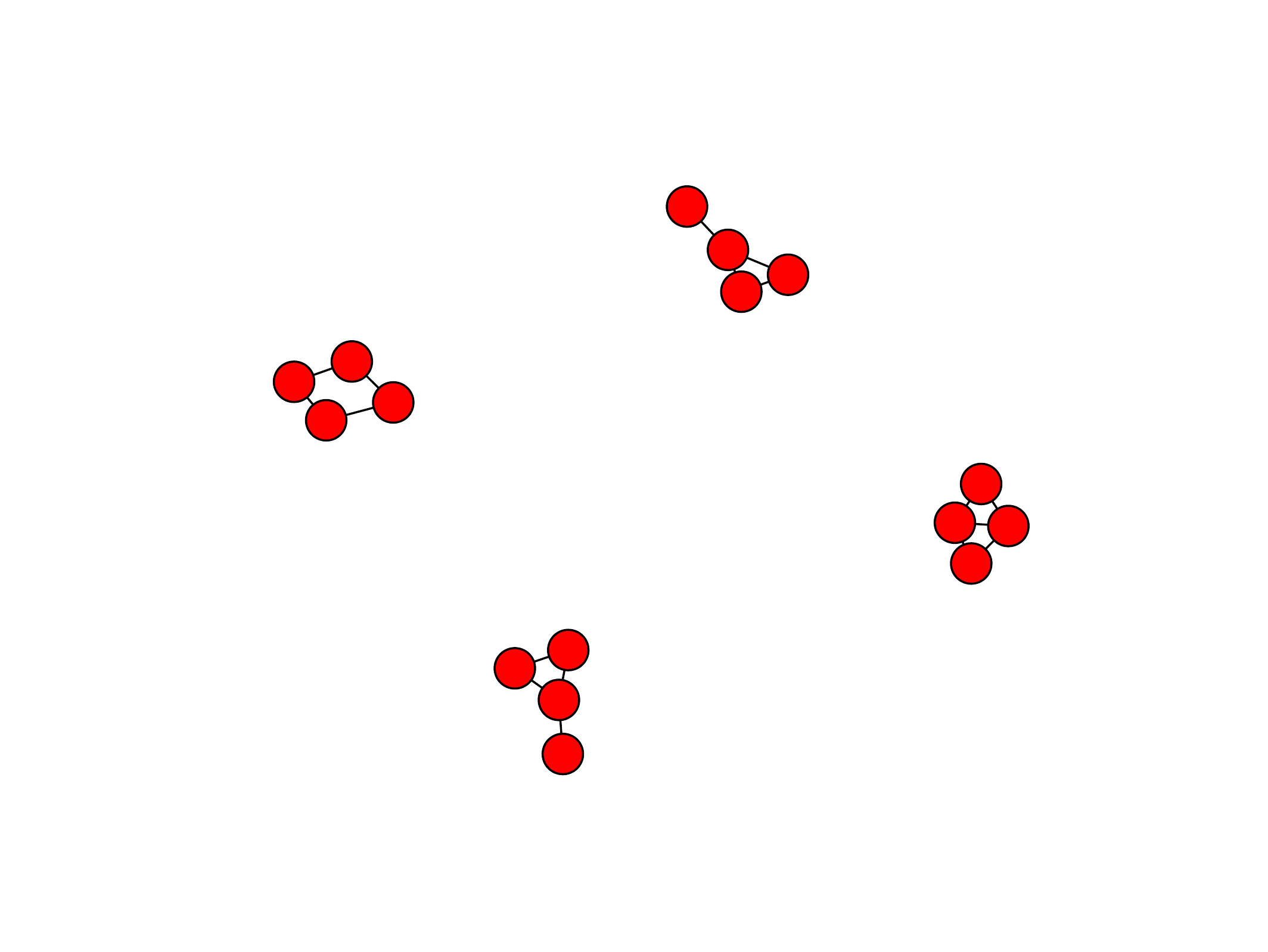}
        \label{2}
    }
    \subfigure[]{
        \includegraphics[width=42mm]{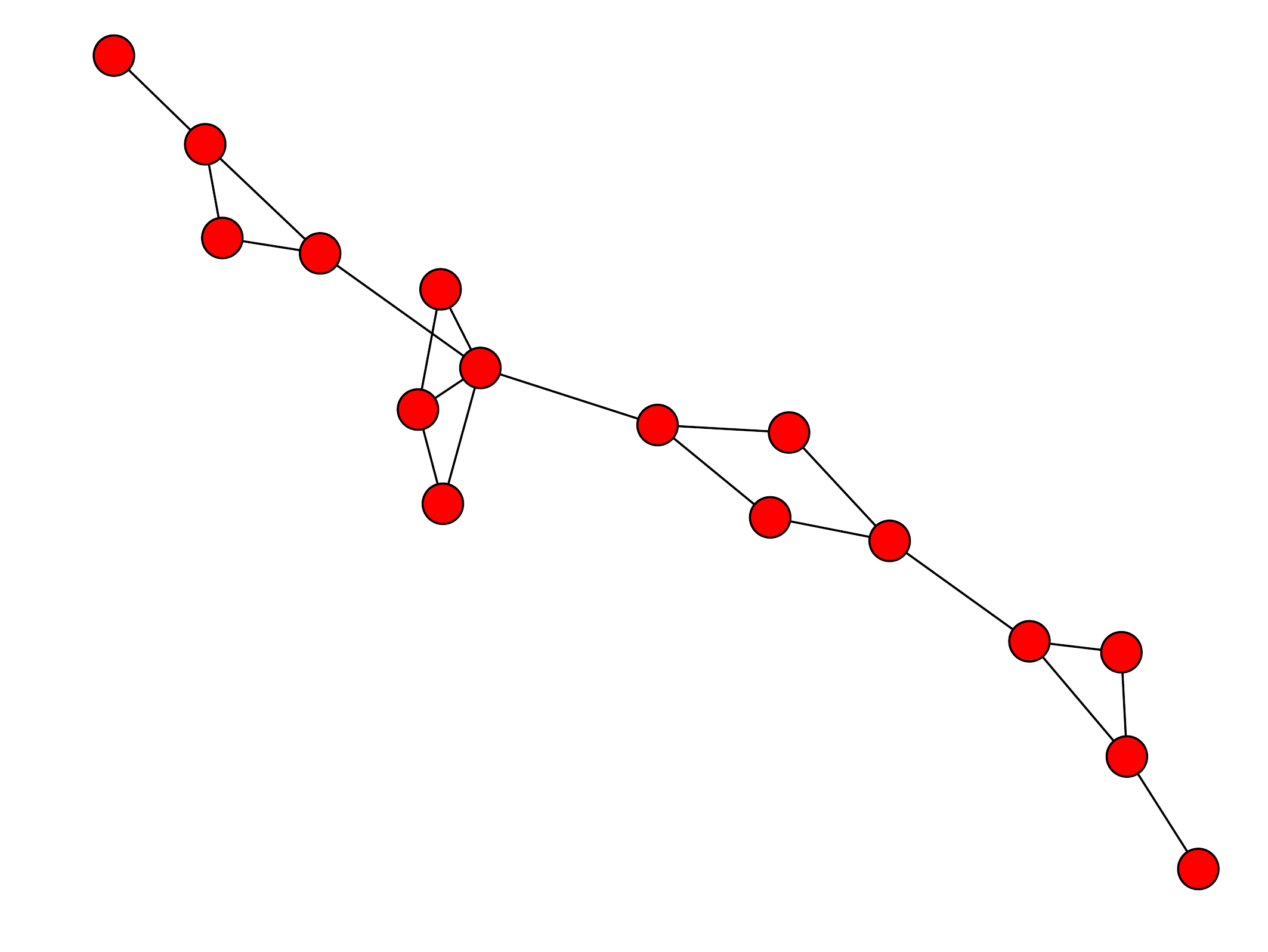}
        \label{3}
    }
    \subfigure[]{
        \includegraphics[width=42mm]{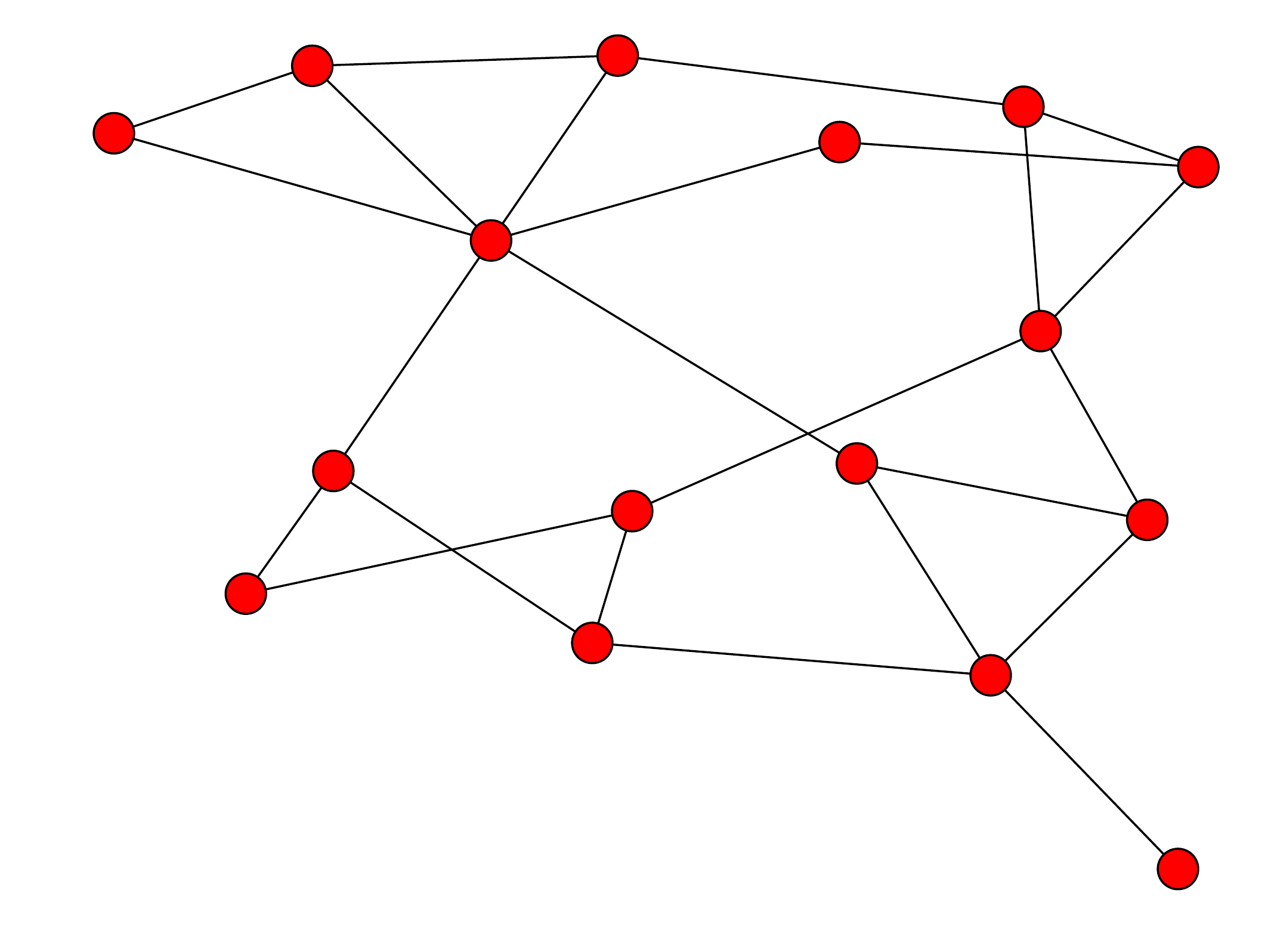}
        \label{4}
    }
\end{center}
\caption{Illustration of some possibilities for the Community Overlap Graph, for a network with 16 communities. Each node represents a community; edges connect pairs of overlapping communities. (a) Non-overlapping communities. (b) Overlapping communities, but clustered, with no path through overlap. (c) Overlapping communities, with only long paths through overlap. (d) Overlapping communities, with short paths solely through community overlap.  All networks illustrated might still be small world, due to the presence of weak ties which do not appear in the community overlap graph. We argue that (d) is the best conceptual model for real world social networks.}
\label{communityOverlapGraph}
\end{figure*}

This leads us to a more general observation:  the `community overlap graph' of any overlapping community network model, which preserves the `small world' feature without random `weak ties', must necessarily  have a low ASPL -- the `community overlap graph' being the graph wherein there is one node per community, and edges connect communities that overlap.
See Figure \ref{communityOverlapGraph} for an illustration of this idea.
A low ASPL in the community overlap graph will typically mean that both simple and complex contagions spread fast; this is thus the crucial feature of models of overlapping community structure, where speed of contagion is concerned.

The low ASPL on real world networks has traditionally been attributed to the existence of `weak ties', rather than a `small world' community overlap graph.
However, we argue that, as in these models, the `small world' can also be explained by overlapping communities; specifically by a low ASPL in the community overlap graph.
This is obviously a matter to be investigated on empirical data.

\subsection{Empirical Networks}
\begin{table*}[ht]
\caption{For the five largest Facebook 100 datasets, we show the size of the Large Connected Component ($LCC$), the size of the LCC formed when weak ties are removed ($LCC_{-WT}$), the average shortest path length (ASPL) within the $LCC_{-WT}$, and the ASPL for the same set of $LCC_{-WT}$ nodes, but with weak ties between these nodes added back in (ASPL $LCC_{+WT}$).}
\label{weakTieRemoval}
\begin{center}
\begin{tabular}{r|rr|r|rr|rr}
\hline
Network & Nodes & Edges     & Weak ties & Size $LCC$ & Size $LCC_{-WT}$ & ASPL $LCC_{-WT}$& ASPL $LCC_{+WT}$ \\
\hline
UPenn  & 41555 & 1362229     &49807  &41536  &39834 &3.10 &3.05 \\
UTexas & 36372 & 1590655     &45727  &36364  &35311 &2.86 &2.84 \\
UF  & 35124 &1465660        &36385 &35111  &34140  &2.90 &2.87 \\
MSU & 32376 & 1118774       &39961 &32361  &31248  &3.02 &2.98 \\
UIllinios  &30810 & 1264428 &31129  &30795  &29878 &2.97 &2.92 \\
\hline
\end{tabular}
\end{center}
\vspace{-12px}
\end{table*}

\subsubsection{Facebook Networks}
While Salath\'{e} and Jones did not investigate the relationship of overlapping structure with epidemic spread, they did conduct empirical work to benchmark an immunization strategy on Facebook data, which is known to be overlapping \cite{lee2010detecting}.
However, they filtered out many of the edges from the Facebook networks, keeping only those edges among users in the same dormitory, or in the same class and college course.
This removes much of the overlap from the community structure, and may neglect relationships formed by shared sporting activities, social interests, friendships, and so on.
We are interested here in a wide range of contagions through this network, such as the spread of viral news or video on the Facebook network, rather than purely whether the users would be in close enough contact to allow disease to spread.
As such, we will perform no filtering, and maintain the original highly-overlapping nature of the networks we study.
Once again, we are concerned with the speed and reach of complex contagions, on these networks.
In particular, we consider the largest 5 publicly available Facebook datasets, extracted from the recently released collection of 100 collegiate social networks, known as the `Facebook 100', of Traud et al. \cite{traud2008community}.

Shi et al. \cite{shi2007networks} previously studied strong ties on two datasets -- one online social network of 2,000 users, and one large subset of the AOL Instant Messenger network, with 140K direct users -- and found that removing weak ties (we refer to ties as `weak' if they are `structurally weak', after Granovetter \cite{granovetter1973strength} ; that is, if the tie is not part of a triangle) from these networks increased the ASPL of the network only slightly.
This idea is of relevance to our research, as we are interested in the speed with which both simple and complex contagions can spread on online social networks; triangles, which are strong ties, are of importance to the spread of complex contagions.
We thus perform a similar analysis on the largest of the `Facebook 100' datasets.
We believe these networks, with their high degree of uptake among the student population, their large size, and considering the now widespread use of the Facebook network, are interesting networks to analyze.

\begin{figure*}[!htb]
\begin{center}
    \subfigure[]{
        \includegraphics[width=55mm]{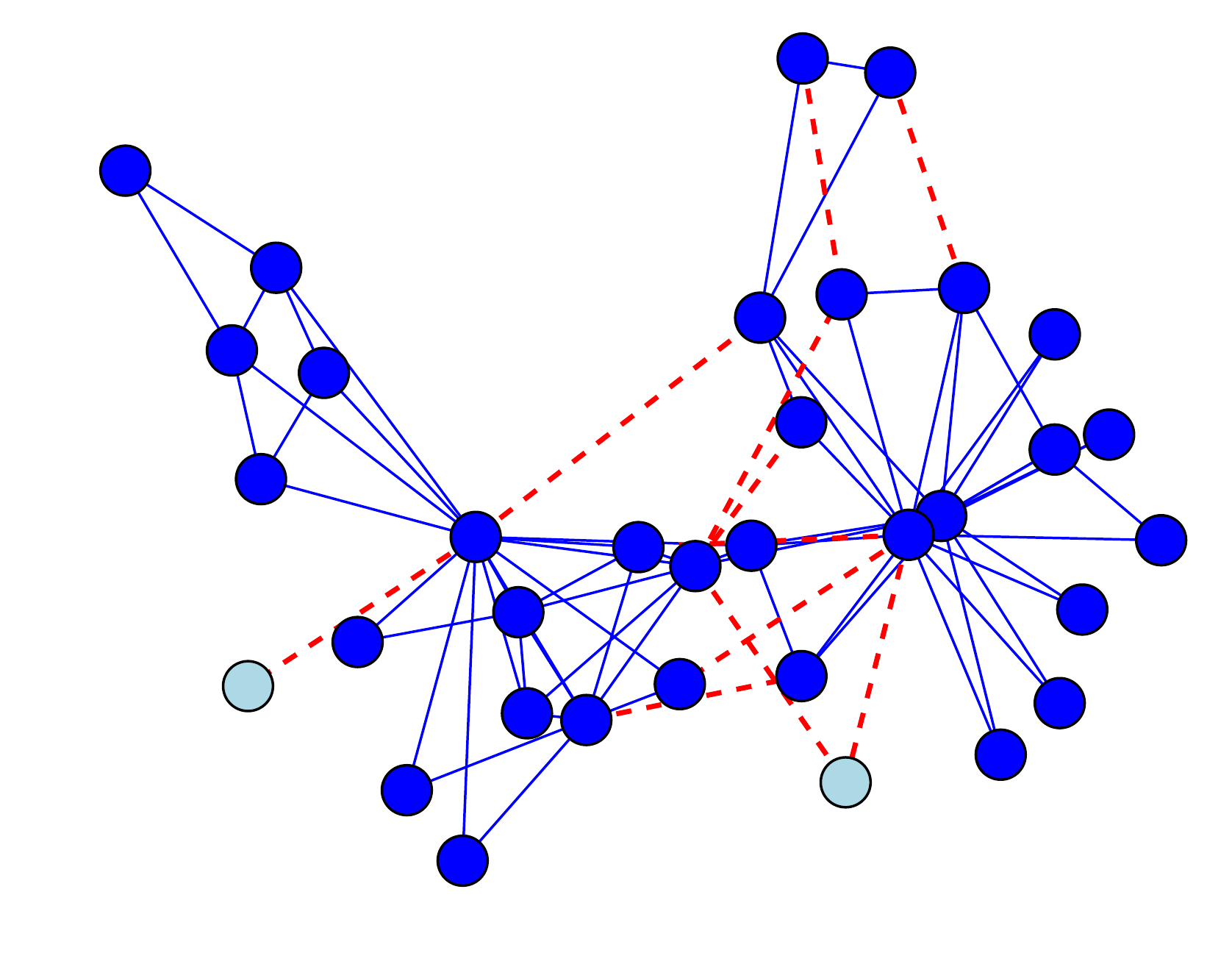}
        \label{1}
    }
    \subfigure[]{
        \includegraphics[width=55mm]{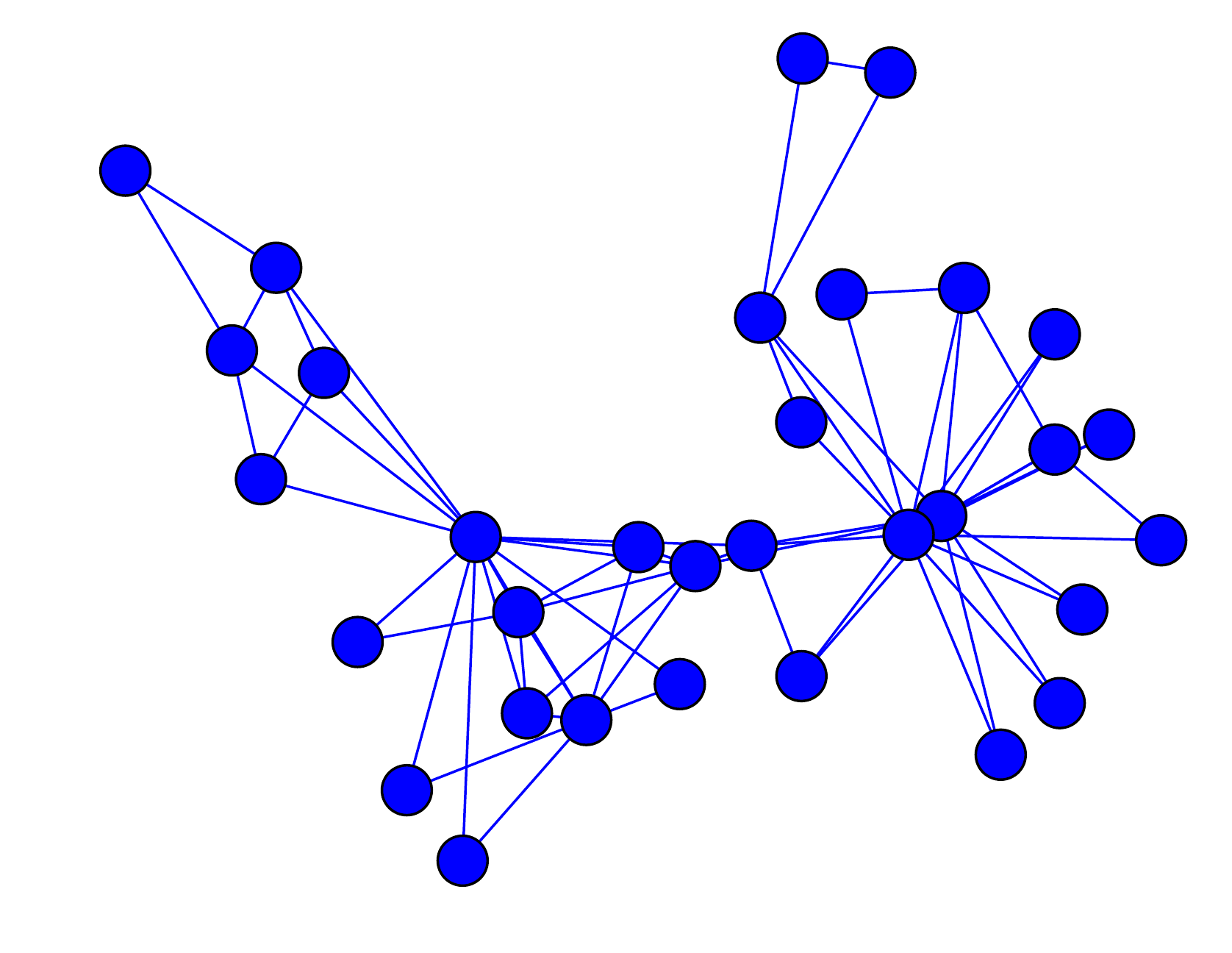}
        \label{2}
    }
    \subfigure[]{
        \includegraphics[width=55mm]{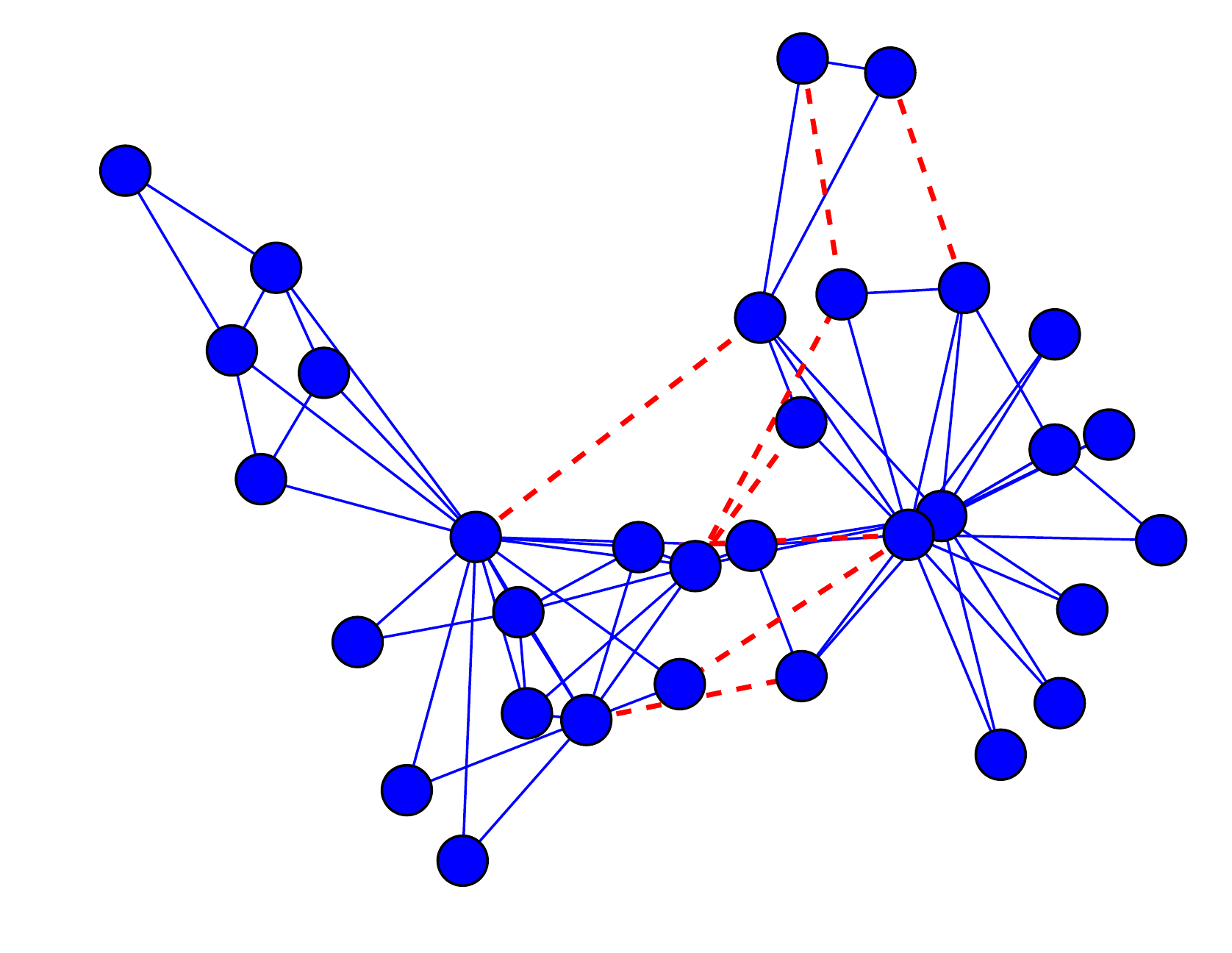}
        \label{3}
    }
\end{center}
\caption{Illustration of experiment to calculate effect of weak ties on ASPL of fixed set of nodes. (a) Original graph, with weak ties shown in red, and nodes only connected by weak ties shown in light blue. (b) LCC of graph connected by strong ties. We call the ASPL among this set of nodes, without weak ties $LCC_{-WT}$ (c) The same set of nodes, but with weak ties between them added back in. We call the ASPL of this graph $LCC_{+WT}$.}
\label{weakTie}
\end{figure*}

Shi et al. study the ASPL on their networks, among all \textit{connected nodes}, as weak ties are removed.
We may be concerned that this is not a fair comparison, as removing some weak ties from the network also reduces the number of connected nodes in the network.
To correct this, we instead perform the following experiment:
First, we remove all weak ties from the network.
We then calculate the largest connected component (LCC) in the remaining network, and also calculate the ASPL among the nodes within this LCC.
We next add back in all weak ties \textit{that are between nodes within the LCC}.
We again calculate the ASPL, among this same set of nodes, and compare results -- Figure \ref{weakTie} illustrates this process.

This method allows us to compare the ASPL among a fixed set of nodes, with and without weak ties.
As some of these networks have a large number of nodes, we use the same ASPL sampling method as \cite{leskovec2008planetary}, sampling the ASPL as calculated to all nodes from each of 1,000 random starting nodes.
We present our results in Table \ref{weakTieRemoval}.  As clearly shown in these results, removing weak ties only disconnects a small amount of non-core nodes from the LCC -- typically less than 5\%.
These results show that on these large social networks adding the weak ties reduces the ASPL only very slightly -- even with no structurally weak ties, the social networks have a very low ASPL, and are `small worlds'.
If our model of networks was of dense non-overlapping community structure, only connected by weak ties, we would have expected a drastic increase in ASPL when we removed the weak ties -- no such increase occurs.
In the influential Watts-Strogatz model of networks, and in the work of Granovetter \cite{granovetter1973strength}, weak ties are very important for fast information flow; however, we see that even if weak ties are removed from online social networks, the low ASPL remains; there is no topological reason why simple contagions should not continue to spread fast, even without the weak ties.
We now proceed to extend this line of argument to consider the paths between cliques, as opposed to simply strong ties.

As motivated earlier, we are interested in the ASPL of the community overlap graph in these networks, in order to examine whether pervasively overlapping community structure could be responsible for the low ASPL.
Ideally, we would generate the set of overlapping communities, calculate their overlap, build the community overlap graph, and calculate the ASPL of it.
However, despite recent work on overlapping community finding algorithms, we remain cautious about interpreting their results.
Specifically, in order for claims about the ASPL in the overlapping community graph to be valid, there must be little chance that the community finding process can \textit{overestimate} the size of the communities it finds.
As such, we will not use a community finding algorithm and we instead will look to maximal cliques -- fully connected subgraphs -- as a conservative and interpretable underestimate of community structure.
Thus, the connectivity of the clique graph -- which we build by adding one node per maximal clique, and edges between cliques that overlap \cite{evans2010clique} -- is a lower bound on the connectivity of the community graph, for the same set of underlying source nodes.
However, the full clique graph in networks of the size we consider, with their many similar maximal cliques, and highly overlapping structure, is computationally infeasible to work with.
As such, we use a proxy to it, as follows:

First, we obtain the network that would be formed by considering, as our communities, maximal cliques of size 5 or greater, that overlap by at least one node.
We calculate this graph by generating the set of all maximal 5 cliques, marking the edges of the source network that are within them, and discarding all other edges.
The ASPL in this graph informs us as to the speed of contagion that flows only using overlapping community structure.
In Table \ref{fiveCliqueOnly} under `5Clique Nodes' we show the number of nodes in the source network that are covered by 5-cliques.
Like with our previous experiment, a large proportion of nodes are covered by 5-cliques, in Facebook.
`5Clique ASPL' shows the ASPL for this set of nodes, which is \textit{short in all cases}.
As the number of the nodes covered by 5-cliques is smaller than in the original graph, for comparison purposes we then generate the graph where all original edges -- i.e. those not in 5-cliques -- between these nodes, are added back in; the value of the ASPL among this same set of nodes is shown under `All ASPL' in the table.
These results show that the networks of these overlapping cliques, which cover the vast majority of the source nodes, have short paths through them; while removing edges not in communities does increase the ASPL, overlapping community structure alone is sufficient to carry a contagion quickly through the network.

\begin{table}[ht]
\caption{For the five largest Facebook 100 datasets, we show ASPL within the Large Connected Component of the network formed when only nodes covered by cliques of size 5 or greater are considered.
We show the number of such nodes for each network (5Clique Nodes), the ASPL for these nodes considering only the paths through the 5-cliques (5Clique ASPL), and the ASPL between these nodes when all other ties between them are added back into the network (All ASPL).}
\label{fiveCliqueOnly}
\begin{center}
\begin{tabular}{r|r|rr}
\hline
Network & 5Clique Nodes &5Clique ASPL &All ASPL\\
\hline
UPenn  & 35940 &3.21 & 2.94\\
UTexas & 32875 &2.94 & 2.79\\
UF  & 31800& 2.96 & 2.81\\
MSU &  28676 &3.13 & 2.90\\
UIllinios  &28034  &3.06 & 2.87 \\
\hline
\end{tabular}
\end{center}
\vspace{-10px}
\end{table}
This means that these empirical networks are `small worlds' in terms of their overlapping community structure.
We argue this makes intuitive sense -- as the Watts Strogatz model shows, only a small portion of random edges is required to decrease the ASPL in a lattice.
Similarly, if only a small portion of overlapping communities are `long-range' in the community overlap graph then the ASPL of the community overlap graph will decrease rapidly.
It is not hard in a modern world to imagine that, in many social networks, and certainly many on-line social networks, while many overlapping communities will be `local', a small proportion of communities are `long-range' in this way; this has deep implications for how we think about diffusion and the role of weak ties on these networks.

\begin{table*}[ht]
\caption{Results of simulation of 1000 complex contagions on each Facebook network. Shown are average time before 50\% nodes were infected, average duration of infections (time before all infectable nodes were infected) and standard deviations $\sigma$. In all cases, $\sigma$ of final infection size was negligible ($<1$).}
\label{facebookComplexContagion}
\begin{center}
\begin{tabular}{r|rr|rr|rr|rr}
\hline
Network & Nodes & Edges     & Avg Time 50\% Infected & $\sigma$ & Avg duration& $\sigma$ &Avg Final Size\\
\hline
UPenn  & 41555 & 1362229     & 3.69 &0.71  &10.51 &0.85 &40443.97 \\
UTexas & 36372 & 1590655     & 3.32 &0.57  &9.52 &0.79 &35689.97  \\
UF  & 35124 &1465660        &3.36 &0.58  &9.57  & 0.79 &34503.95 \\
MSU & 32376 & 1118774       &3.61 &0.66  &10.21  &0.85 &31635.97 \\
UIllinios  &30810 & 1264428 &3.48  &0.64  &10.04  & 0.8 &30196.96 \\
\hline
\end{tabular}
\end{center}
\end{table*}

Having established that these networks are made `small world' by their overlapping community structure, we now conduct simulation of complex contagions, on these same networks, and present our results in Table \ref{facebookComplexContagion}.
We find, as expected given the short paths through overlapping community structure, and the speed with which complex contagions move through overlapping communities, that the complex contagions diffuse fast throughout these large networks.
The average time for a complex contagion to reach 50\% of the nodes in all networks is less than 4 time steps.
In each case the final size of the contagion -- the total number of nodes it reaches -- is close to the total number of nodes in the network.
The deviation of this number, on each network, is negligible -- in each simulation practically all nodes which a complex contagion can reach, are reached, and the times to reach all these nodes is about 10 time steps.
\subsubsection{Twitter}
Finally, we consider diffusion on a much larger class of social graph -- that of Twitter, with tens of millions of users, and global reach.
This is a very different network to Facebook, in that user relationships are directed `follower' relations, rather than bidirectional friend relations.
It is also topical -- much recent discussion has focused on the flow of information through Twitter; it has been theorized that certain contagions, such as the spread of participation in a revolution, are complex, requiring multiple reaffirmations before they will be adopted -- and so the potential speed of complex contagions on networks like Twitter is of current interest.

We simulate complex contagions on the entire Twitter network as of mid 2009, as gathered by Kwak et al. \cite{kwak2010twitter}, and present results on the speed of such contagions.
We first conduct simulations of complex contagions on the raw Twitter network.
This network has over 41M nodes, and almost 1.4B edges.  
However, Twitter is a directed network, and many users use it as a news service, where they passively follow others, but are not densely connected.
For the study of contagions on Twitter, we are primarily interested in users in the core of the network.
We calculate the strongly connected components (SCC) of this graph, and find a large SCC with approximately 33.5M nodes in it, and no other non-trivial SCCs.
Of these users in the SCC, we find that approximately 5.8M of them are following only a single other user, and are thus not potentially infectable by a complex contagion.
This leaves approximately 27.7M users reachable by a complex contagion, who are in the densely connected core of the network.
We find that repeat complex contagion simulations either infect ~27.04M users, or only a very small number of users (fewer than 100).

\begin{figure}[!htb]
    \includegraphics[width=90mm]{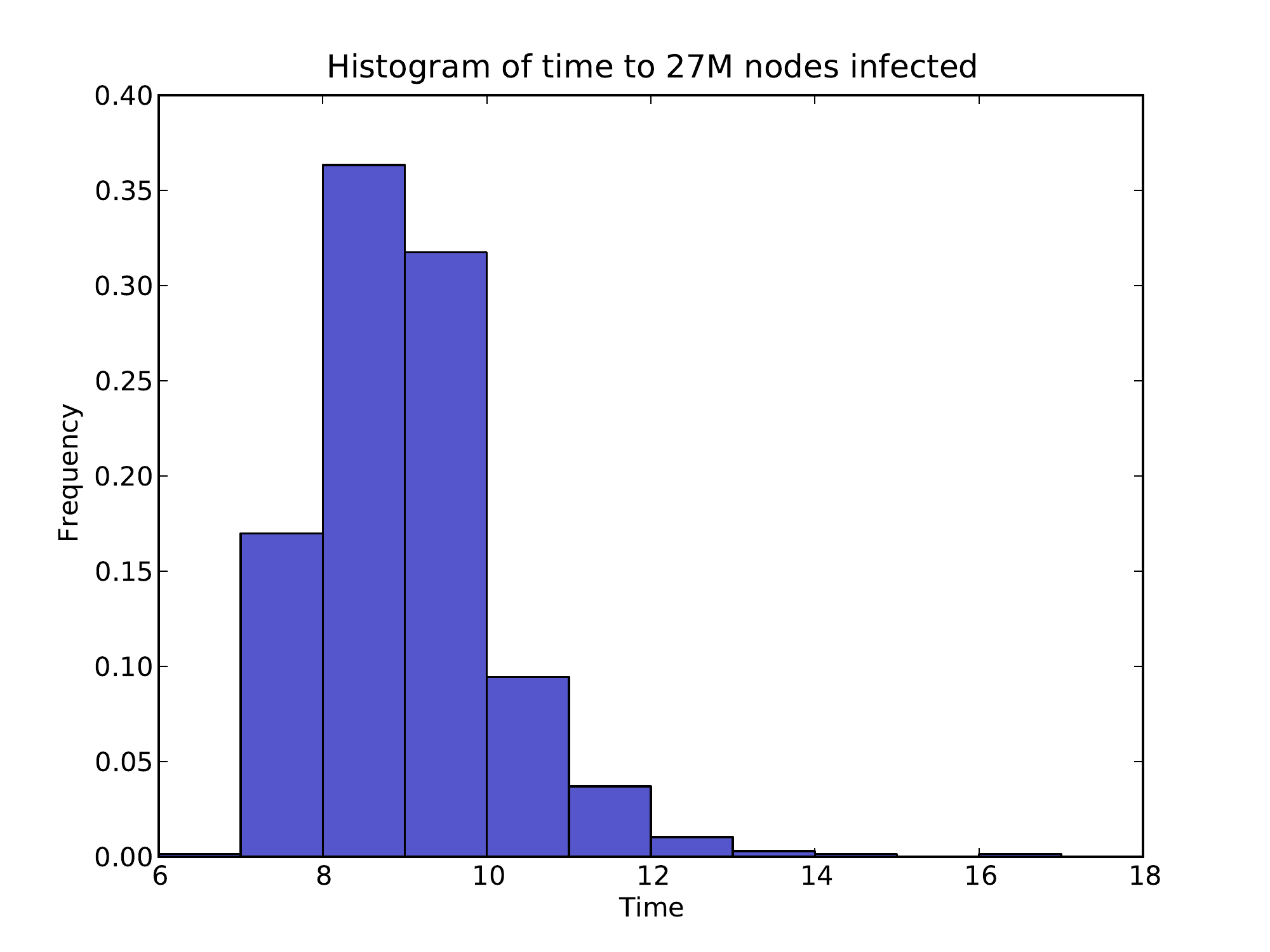}
\caption{Complex contagion in Twitter; shown is the distribution of time to 27M nodes Infected, for 677 simulated contagions which became established.}
\label{timeForComplexContagionTo27MTwitterSCC}
\vspace{-4px}
\end{figure}

We are interested in the possible speed of such a complex contagion over the Twitter core.
There are always a very small number of nodes that the contagion takes a long time to reach -- users not well connected to the rest of the core.
We are interested in how fast a complex contagion can move through the core of the Twitter network, and thus present the distribution of the number of simulation timesteps it takes a complex contagion to reach 27M nodes, in Figure \ref{timeForComplexContagionTo27MTwitterSCC}.
Like previous work, we only show results for those contagions that propagated beyond their initial neighbours -- in practice, any contagion that reached over 100 nodes reached all 27M.
We note that if a contagion has already reached over 1,000 nodes, it then only takes an average of 5.11 time steps (standard deviation 0.34) to reach 27M nodes.
It is clear that once a complex contagion is established on Twitter, it may diffuse rapidly to tens of millions of users, despite its complexity.

Our key result here is that the vast majority of nodes in the core are typically reached by the complex contagion in a small number of timesteps.
This shows that the core of Twitter is a `small-world' where complex contagions are concerned; it is possible for a complex contagion to flow very quickly through it.
This is not a result we would \textit{a priori} expect: if we thought the network to be made a small world by random weak ties, rather than overlapping community structure, we would have expected the complex contagion to diffuse slowly.

\section{Conclusion}
If we think of community structure as easily partitionable and connected only by near-random weak ties, then increasing community structure slows the spread of the contagions, as shown by \cite{salathe2010dynamics}.
However, it has previously been established that in many large online social networks, if we define community as having few external edges, then such structures do not exist in the cores of large social networks \cite{leskovec2008statistical}, and that on many networks, where overlapping community structure does exist, these communities are not separable from each other by partitioning; i.e. they overlap pervasively \cite{reid2011partitioning}.

In light of this, we have modified the Salath\'{e} and Jones model to allow us to investigate diffusion in network models with overlapping community structure.
We show that increasing overlapping community structure does not have the slowing effect on simple contagions which increasing non-overlapping community structure does.
We also find that complex contagions, which it is theorized \cite{centola2007complex} require wide bridges to spread, actually spread faster as community overlap increases, in such models.

It has long been thought that `structurally weak' ties have been responsible for the small world effect in social networks, and are particularly topologically important for diffusion.
However, in line with some previous work \cite{shi2007networks}, we have shown that short paths exist through popular social networks through strong ties alone.
We have also considered 5-cliques as an underestimate of community structure, and shown that short paths exist through the network of overlapping cliques.
Further, we have studied the spread of complex contagions using simulation on large empirical networks, and found that complex contagions can spread fast on these topologies.

We conclude that the role of structurally weak ties in diffusion is over-stated; as applied to online social networks, a conceptual model of the world in which communities overlap pervasively also explains the `small-world' phenomenon, in an information diffusion context.
This has implications for how we think about the speed with which an online social network can carry both simple and complex contagions, and which ties and structure are topologically important to diffusion processes.
These results should inform the development of algorithmic approaches to mine sets of nodes and community structure crucial to diffusion processes.

\bibliographystyle{IEEEtran}

\bibliography{IEEEabrv,mybib}

\begin{thebibliography}{10}
\providecommand{\url}[1]{#1}
\csname url@samestyle\endcsname
\providecommand{\newblock}{\relax}
\providecommand{\bibinfo}[2]{#2}
\providecommand{\BIBentrySTDinterwordspacing}{\spaceskip=0pt\relax}
\providecommand{\BIBentryALTinterwordstretchfactor}{4}
\providecommand{\BIBentryALTinterwordspacing}{\spaceskip=\fontdimen2\font plus
\BIBentryALTinterwordstretchfactor\fontdimen3\font minus
  \fontdimen4\font\relax}
\providecommand{\BIBforeignlanguage}[2]{{%
\expandafter\ifx\csname l@#1\endcsname\relax
\typeout{** WARNING: IEEEtran.bst: No hyphenation pattern has been}%
\typeout{** loaded for the language `#1'. Using the pattern for}%
\typeout{** the default language instead.}%
\else
\language=\csname l@#1\endcsname
\fi
#2}}
\providecommand{\BIBdecl}{\relax}
\BIBdecl

\bibitem{milgram1967small}
S.~Milgram, ``{The small world problem},'' \emph{Psychology today}, vol.~2,
  no.~1, pp. 60--67, 1967.

\bibitem{watts1998collective}
D.~Watts and S.~Strogatz, ``{Collective dynamics of 'small-world' networks},''
  \emph{Nature}, vol. 393, no. 6684, pp. 440--442, 1998.

\bibitem{aral1999sexual}
S.~Aral, ``{Sexual network patterns as determinants of STD rates: paradigm
  shift in the behavioral epidemiology of STDs made visible},'' \emph{Sexually
  transmitted diseases}, vol.~26, no.~5, p. 262, 1999.

\bibitem{helleringer2007sexual}
S.~Helleringer and H.~Kohler, ``{Sexual network structure and the spread of HIV
  in Africa: evidence from Likoma Island, Malawi},'' \emph{Aids}, vol.~21,
  no.~17, p. 2323, 2007.

\bibitem{pastor2001epidemic}
R.~Pastor-Satorras and A.~Vespignani, ``{Epidemic spreading in scale-free
  networks},'' \emph{Physical review letters}, vol.~86, no.~14, pp. 3200--3203,
  2001.

\bibitem{Newman2002a}
M.~Newman, ``{Spread of epidemic disease on networks},'' \emph{Physical Review
  E}, vol.~66, no.~1, pp. 1--11, Jul. 2002.

\bibitem{Meloni2009}
S.~Meloni, A.~Arenas, and Y.~Moreno, ``{Traffic-driven epidemic spreading in
  finite-size scale-free networks.}'' \emph{PNAS}, vol. 106, no.~40, pp.
  16\,897--902, Oct. 2009.

\bibitem{fortunato2009community}
S.~Fortunato, ``{Community detection in graphs},'' \emph{Physics Reports},
  2009.

\bibitem{guillaume2006bipartite}
J.~Guillaume and M.~Latapy, ``{Bipartite graphs as models of complex
  networks},'' \emph{Physica A: Statistical and Theoretical Physics}, vol. 371,
  no.~2, pp. 795--813, 2006.

\bibitem{botha2010community}
L.~Botha and S.~Kroon, ``{A Community-Based Model of Online Social Networks},''
  \emph{SNA-KDD}, 2010.

\bibitem{newman2001random}
M.~Newman, S.~Strogatz, and D.~Watts, ``{Random graphs with arbitrary degree
  distributions and their applications},'' \emph{Physical Review E}, vol.~64,
  no.~2, p. 026118, 2001.

\bibitem{lancichinetti2009benchmarks}
A.~Lancichinetti and S.~Fortunato, ``{Benchmarks for testing community
  detection algorithms on directed and weighted graphs with overlapping
  communities},'' \emph{Physical Review E}, vol.~80, no.~1, p. 16118, 2009.

\bibitem{liu2005epidemic}
Z.~Liu and B.~Hu, ``{Epidemic spreading in community networks},'' \emph{EPL
  (Europhysics Letters)}, vol.~72, p. 315, 2005.

\bibitem{chu2009epidemic}
X.~Chu, J.~Guan, Z.~Zhang, and S.~Zhou, ``{Epidemic spreading in weighted
  scale-free networks with community structure},'' \emph{Journal of Statistical
  Mechanics: Theory and Experiment}, vol. 2009, p. P07043, 2009.

\bibitem{salathe2010dynamics}
M.~Salath{\'e} and J.~Jones, ``{Dynamics and control of diseases in networks
  with community structure},'' \emph{PLoS Comput Biol}, vol.~6, no.~4, p.
  e1000736, 2010.

\bibitem{leskovec2008statistical}
J.~Leskovec, K.~Lang, A.~Dasgupta, and M.~Mahoney, ``{Statistical properties of
  community structure in large social and information networks},'' in
  \emph{Proceeding of the 17th international conference on World Wide
  Web}.\hskip 1em plus 0.5em minus 0.4em\relax ACM, 2008, pp. 695--704.

\bibitem{reid2011partitioning}
F.~Reid, A.~McDaid, and N.~Hurley, ``{Partitioning Breaks Communities},''
  \emph{2011 Advances in Social Network Analysis and Mining}, 2011.

\bibitem{centola2007complex}
D.~Centola and M.~Macy, ``{Complex contagions and the weakness of long ties},''
  \emph{American Journal of Sociology}, vol. 113, no.~3, pp. 702--34, 2007.

\bibitem{Centola2007a}
D.~Centola, V.~Eguiluz, and M.~Macy, ``{Cascade dynamics of complex
  propagation},'' \emph{Physica A: Statistical Mechanics and its Applications},
  vol. 374, no.~1, pp. 449--456, Jan. 2007.

\bibitem{romero2011differences}
D.~Romero, B.~Meeder, and J.~Kleinberg, ``{Differences in the mechanics of
  information diffusion across topics: idioms, political hashtags, and complex
  contagion on twitter},'' in \emph{Proceedings of the 20th international
  conference on World wide web}.\hskip 1em plus 0.5em minus 0.4em\relax ACM,
  2011, pp. 695--704.

\bibitem{granovetter1973strength}
M.~Granovetter, ``{The strength of weak ties},'' \emph{ajs}, vol.~78, no.~6, p.
  1360, 1973.

\bibitem{shi2007networks}
X.~Shi, L.~Adamic, and M.~Strauss, ``{Networks of strong ties},'' \emph{Physica
  A: Statistical Mechanics and its Applications}, vol. 378, no.~1, pp. 33--47,
  2007.

\bibitem{van2006strong}
M.~van~der Leij and S.~Goyal, \emph{{Strong ties in a small world}}.\hskip 1em
  plus 0.5em minus 0.4em\relax Tinbergen Institute, 2006.

\bibitem{ahn2010link}
Y.~Ahn, J.~Bagrow, and S.~Lehmann, ``{Link communities reveal multiscale
  complexity in networks},'' \emph{Nature}, vol. 466, no. 7307, pp. 761--764,
  2010.

\bibitem{evans2009line}
T.~Evans and R.~Lambiotte, ``{Line graphs, link partitions, and overlapping
  communities},'' \emph{Physical Review E}, vol.~80, no.~1, p. 016105, 2009.

\bibitem{anderson1979population}
R.~Anderson and R.~May, ``{Population biology of infectious diseases: Part
  I},'' \emph{Nature}, vol. 280, no. 5721, pp. 361--367, 1979.

\bibitem{kermack1150}
W.~Kermack, ``{0. \& McKendrick, AG 1927 A contribution to the mathematical
  theory of epidemics},'' in \emph{Proc. R. Soc. Lond. A}, vol. 115, pp.
  700--721.

\bibitem{good2010performance}
B.~Good, Y.~De~Montjoye, and A.~Clauset, ``{Performance of modularity
  maximization in practical contexts},'' \emph{Physical Review E}, vol.~81,
  no.~4, p. 046106, 2010.

\bibitem{lee2010detecting}
C.~Lee, F.~Reid, A.~McDaid, and N.~Hurley, ``{Detecting highly overlapping
  community structure by greedy clique expansion},'' \emph{KDD SNA 2010}, 2010.

\bibitem{traud2008community}
A.~Traud, E.~Kelsic, P.~Mucha, and M.~Porter, ``{Community structure in online
  collegiate social networks},'' \emph{arXiv}, vol. 809, 2008.

\bibitem{leskovec2008planetary}
J.~Leskovec and E.~Horvitz, ``{Planetary-scale views on a large
  instant-messaging network},'' in \emph{Proceeding of the 17th international
  conference on World Wide Web}.\hskip 1em plus 0.5em minus 0.4em\relax ACM,
  2008, pp. 915--924.

\bibitem{evans2010clique}
T.~Evans, ``Clique graphs and overlapping communities,'' \emph{Journal of
  Statistical Mechanics: Theory and Experiment}, vol. 2010, p. P12037, 2010.

\bibitem{kwak2010twitter}
H.~Kwak, C.~Lee, H.~Park, and S.~Moon, ``{What is Twitter, a social network or
  a news media?}'' in \emph{Proceedings of the 19th international conference on
  World wide web}.\hskip 1em plus 0.5em minus 0.4em\relax ACM, 2010, pp.
  591--600.

\end{thebibliography}

\end{document}